\documentclass[12pt,preprint]{aastex63}
\usepackage{amssymb, amsmath}
\usepackage{color}
\usepackage{morefloats}
\usepackage{hyperref}

\begin{document}

\title{The Red Supergiant Binary Fraction of the \\
Large Magellanic Cloud}

\correspondingauthor{kneugent@uw.edu}

\author[0000-0002-5787-138X]{Kathryn F.\ Neugent}
\affiliation{Department of Astronomy, University of Washington, Seattle, WA, 98195}
\affiliation{Lowell Observatory, 1400 W Mars Hill Road, Flagstaff, AZ 86001}

\author[0000-0003-2184-1581]{Emily M.\ Levesque}
\affiliation{Department of Astronomy, University of Washington, Seattle, WA, 98195}

\author[0000-0001-6563-7828]{Philip Massey}
\affiliation{Lowell Observatory, 1400 W Mars Hill Road, Flagstaff, AZ 86001}
\affiliation{Department of Astronomy and Planetary Science, Northern Arizona University, Flagstaff, AZ, 86011-6010}

\author[0000-0003-2535-3091]{Nidia I.\ Morrell}
\affiliation{Las Campanas Observatory, Carnegie Observatories, Casilla 601, La Serena, Chile}

\author[0000-0001-7081-0082]{Maria R.\ Drout}
\affiliation{Department of Astronomy and Astrophysics, University of Toronto, 50 St. George St., Toronto, ON M5S 3H4, Canada}
\affiliation{The Observatories of the Carnegie Institution for Science, 813 Santa Barbara St., Pasadena, CA 91101, USA}

\begin{abstract}
The binary fraction of unevolved massive stars is thought to be 70-100\% but there are few observational constraints on the binary fraction of the evolved version of a subset of these stars, the red supergiants (RSGs). Here we identify a complete sample of RSGs in the Large Magellanic Cloud (LMC) using new spectroscopic observations and archival UV, IR and broadband optical photometry. We find 4090 RSGs with $\log L/L_\odot > 3.5$ with 1820 of them having $\log L/L_\odot > 4$, which we believe is our completeness limit. We additionally spectroscopically confirmed 38 new RSG+B star binaries in the LMC, bringing the total known up to 55. We then estimated the binary fraction using a k-nearest neighbors algorithm that classifies stars as single or binary based on photometry with a spectroscopic sample as a training set. We take into account observational biases such as line-of-sight stars and binaries in eclipse while also calculating model-dependent corrections for RSGs with companions that our observations were not designed to detect. Based on our data, we find an initial result of $13.5^{+7.56}_{-6.67}\%$ for RSGs with O or B-type companions. Using the Binary Population and Spectral Synthesis (BPASS) models to correct for unobserved systems, this corresponds to a total RSG binary fraction of $19.5^{+7.6}_{-6.7}\%$. This number is in broad agreement with what we would expect given an initial OB binary distribution of 70\%, a predicted merger fraction of 20-30\% and a binary interaction fraction of 40-50\%.
\end{abstract}

\section{Introduction}
Red supergiants (RSGs) are the evolved descendants of 8-30$M_{\odot}$ OB main sequence stars. After these luminous, hot stars burn through their core hydrogen, they evolve off the main-sequence and briefly (a couple hundred thousand years) pass through the yellow supergiant (YSG) phase before cooling down to temperatures below $T_{\rm eff} = 4300$K and drastically expanding in radius to reach sizes hundreds or even thousands of times larger than the radius of the Sun. The vast majority of these stars then end their lives as Type II-P supernovae, though some higher mass RSGs evolve back bluewards to higher temperatures prior to core-collapse (e.g. \citealt{Ekstrom2012}).

While RSGs have been a topic of great interest for decades, until the past few years not much was known about their binary properties. As recently as 2018, only around a dozen confirmed binary RSGs were known (see references in \citealt{NeugentRSG}), all of them being in our own Galaxy. This low binary fraction of RSGs is in contrast with the relatively high binary fraction of their unevolved counterparts -- the OB stars. While the binary fraction of O stars is still contested, the ``corrected" binary fraction of OB stars is $\sim50-60\%$ \citep{Sana30Dor, Dunstall2015}, with some evidence it could be much higher \citep{Gies08, SanaSci}. What then happens to all of the binaries once the more massive star evolves into a RSG? In some close systems the binaries will experience Roche lobe overflow (RLOF). When this occurs, it can either lead to stable mass transfer, where the primary will be stripped and will thus never evolve into a RSG (see discussion in \citealt{Trevor2018}), or unstable mass transfer, where the mass is accreted by the secondary faster than it can absorb. In this second case of unstable mass transfer, the secondary will also overflow its Roche lobe and the two stars will enter the common envelope phase. At this point, depending on their proximity, the primary will still evolve into a RSG and overtime could merge with its less massive companion (an overview of these scenarios is further described in Ch.\ 5 of \citealt{LevesqueRSGs}). However, in the systems with large enough separations, the binary companion should remain.

\citet{NeugentRSG,NeugentRSGII} began investigating the binary properties of RSGs, first by determining what types of stars should exist in a binary system with a RSG from an evolutionary point of view. According to \citet{Ekstrom2012}, the least massive unevolved star that will turn into a RSG (a B-type star with an initial mass of $8M_\odot$) turns off the main sequence at 7.6Myr. If we then look at \citet{Bern96}, we find that the contraction time (or time to the zero age main sequence; ZAMS) for a $3M_\odot$ star is 7.2Myr. Thus, any star less than $3M_\odot$ will not have formed by the time an $8M_\odot$ star has reached the RSG phase and will still be a proto-star. Relating this back to the spectral type of the RSG companions, a $3M_\odot$ on the main sequence is approximately an A0V. Thus, anything more massive (i.e., B-type stars with a few O-type stars) will be the companions to the RSGs. Hotter companions such as Wolf-Rayets (WRs) could theoretically exist in systems with RSGs but \citet{NeugentRSG} found such situations were extremely rare. Additionally, this is what is seen observationally -- all of the dozen known RSG binaries are in systems with B-type companions (see Table 1 in \citealt{NeugentRSGII}). 

To find more, \citet{NeugentRSG,NeugentRSGII} devised a set of photometric criteria to identify RSG+B star binaries using readily available archival data and set off in search of spectroscopic confirmation. After observing a large set of RSG+B star binary candidates in our Local Group galaxies, we have now spectroscopically confirmed 251 new RSG+B star binaries over the last two years - 22 in the SMC, 47 in the LMC (both described in this paper), 88 in M31, and 94 in M33 (some discussed in \citealt{NeugentRSGII}, others to be described in future work). This is a factor of 20 increase over the previously known number of RSG binaries when we started our search with \citet{NeugentRSG}.

At this point we are able to place direct constraints on the RSG binary fraction in one of the galaxies we have surveyed, the Large Magellanic Cloud (LMC). We've chosen to focus our efforts on this one galaxy for several reasons: excellent near infrared (NIR) photometry from 2MASS combined with proper motion estimates from {\it Gaia} allows us to identify a complete sample of RSGs within the galaxy down to a reasonable luminosity cutoff of $\log L/L_\odot = 4$; the LMC is well covered with both {\it GALEX} and the $U,B,V,I$ photometric catalog of \citet{ZaritskyLMC} and the resulting near ultraviolet (NUV) and optical colors allow us to identify possible B star companions; it has a well known and understood metallicity (unlike much of M31 and M33); and we've completed several extensive observing runs spectroscopically confirming RSG+B star binaries over a wide range of color-color space such that we understand our completeness rates. Here we provide a first look at the binary fraction of RSGs at the sub-solar metallicity of the LMC.

Our survey was designed to primarily be sensitive to RSG+OB star companions (though we expect to find few O stars due to their short lifetimes) given the reasoning discussed above. We then rely on the Binary Population and Spectral Synthesis (BPASS) models (v2.2.1; \citealt{BPASS1,BPASS2}) to estimate the model-dependent, but small, correction factors to the RSG binary fraction which we are not observationally sensitive. This method is purposefully not sensitive to RSG+protostars and additionally lacks sensitivity to RSGs in systems with other RSGs and the even shorter lived YSGs; however, based on evolutionary timescales, these pairings should be rare.

These results present the first galaxy-wide and complete study of the binary fraction of RSGs and can be used to compare with evolutionary and population synthesis models. The recent result that $\sim$60\% of massive stars may \emph{interact} with their binary companions throughout their lives \citep{SanaSci} has a profound impact on the predicted populations of supernovae \citep{Eldridge2018,Zapartas2019}, gravitational wave sources \citep[e.g.][]{Tauris2017}, and the ionizing radiation from stellar populations \citep{Gotberg2019,Stanway2016}. However, the details of these predictions depend not only on the initial binary conditions, but also on the outcomes of simplified prescriptions for parameters such as the mass transfer efficiency and outcomes of common envelope evolution \cite[e.g.][]{Podsiadlowski1992,Wellstein1999,Eldridge2008}. As the OB binary fraction and properties becomes better established, this measurement of the binary fraction of their more evolved descendent stars will provide an important boundary condition to test our models of binary evolution.

To calculate the binary fraction of RSGs in the LMC, we first identified a complete sample of RSGs in the LMC photometrically using 2MASS NIR colors and {\it Gaia} to confirm membership. This process is described in Section 2. We then selected a subset of these stars to spectroscopically confirm as RSG+B binaries, as detailed in Section 3. In Section 4 we discuss how we calculated the final binary fraction, including our errors, and in Section 5 we place this in context with the observed binary fraction for other types of massive stars while also comparing our results to model predictions. Finally, we conclude in Section 6. The accompanying Appendix describes how we measured the physical properties of the spectroscopically confirmed RSGs. 

\section{Identifying Red Supergiants}
To calculate the binary fraction of RSGs in the LMC, we first needed to identify a parent sample of all LMC RSGs, aiming to be as complete as possible in order to make the statistics robust. We will then later determine what fraction of these have binary companions. Our goal was to select a complete sample of RSGs down to $\log L/L_\odot = 4$ which corresponds to a minimum initial mass of around $9M_\odot$ (see Fig. 2 in \citealt{Ekstrom2012}). Such a sample will be contaminated both by Galactic foreground stars (nearby red dwarfs) and by the brighter asymptotic giant branch (AGB) stars in the LMC. We eliminated these using the same procedure that we recently used for M31 RSGs \citep{UKIRT}: foreground stars were removed using {\it Gaia} data, and AGB stars were separated from RSGs using cuts in a ($J-K_s$, $K_s$) color-magnitude diagram (CMD), following the pioneering work of \citet{Yang2019}. In order to select a sample of RSGs that was unbiased by the presence of a hot companion, we chose to rely upon 2MASS $J$ and $K$ photometry as the near-IR (NIR) colors will be relatively insensitive to the presence of a hot companion. Thus, our selection criteria will allow us to determine a complete sample of all RSGs, binary and non-binary alike.

\subsection{Selecting Red Stars From 2MASS}
We began by selecting sources from the 2MASS point-source catalog \citep{2MASS} within 210\arcmin\ of the center of the LMC, taken to be $\alpha_{\rm J2000}$=05:18:00 and $\delta_{\rm J2000}$=-68:45:00, chosen to match the same field used for the recent survey for WR stars in the LMC \citep{MasseyMCWRI,FinalCensus} and encompassing the entire optical disk of the galaxy. We kept only objects with the best 2MASS photometry, i.e., with quality flags of ``AAA," and ``artifact contamination" flags of ``000."  Our initial selection was restricted to stars with $K_s\le 13$, and $J-K_s\geq 0.5$. This left us with a sample of 87,637 stars. 

These magnitude and color limits were chosen to be extremely generous. A $K_s=13$ star in the LMC would have $\log L/L_\odot \sim 3.0-3.3$, adopting a distance to the LMC of 50.0~kpc, and the equations given in Table 4 of \cite{UKIRT}.  This is much smaller than our completeness goal of $\log L/L_\odot \sim$ 4.0. Similarly, a lightly reddened $J-K_s=0.5$ star will have an effective temperature ($T_{\rm{eff}}$) of 5000~K (using equation 1 from \citealt{NeugentLMC}), much warmer than the $\sim4200$~K upper temperature limit for RSGs we will employ below.

\begin{figure}
\includegraphics[width=0.5\textwidth]{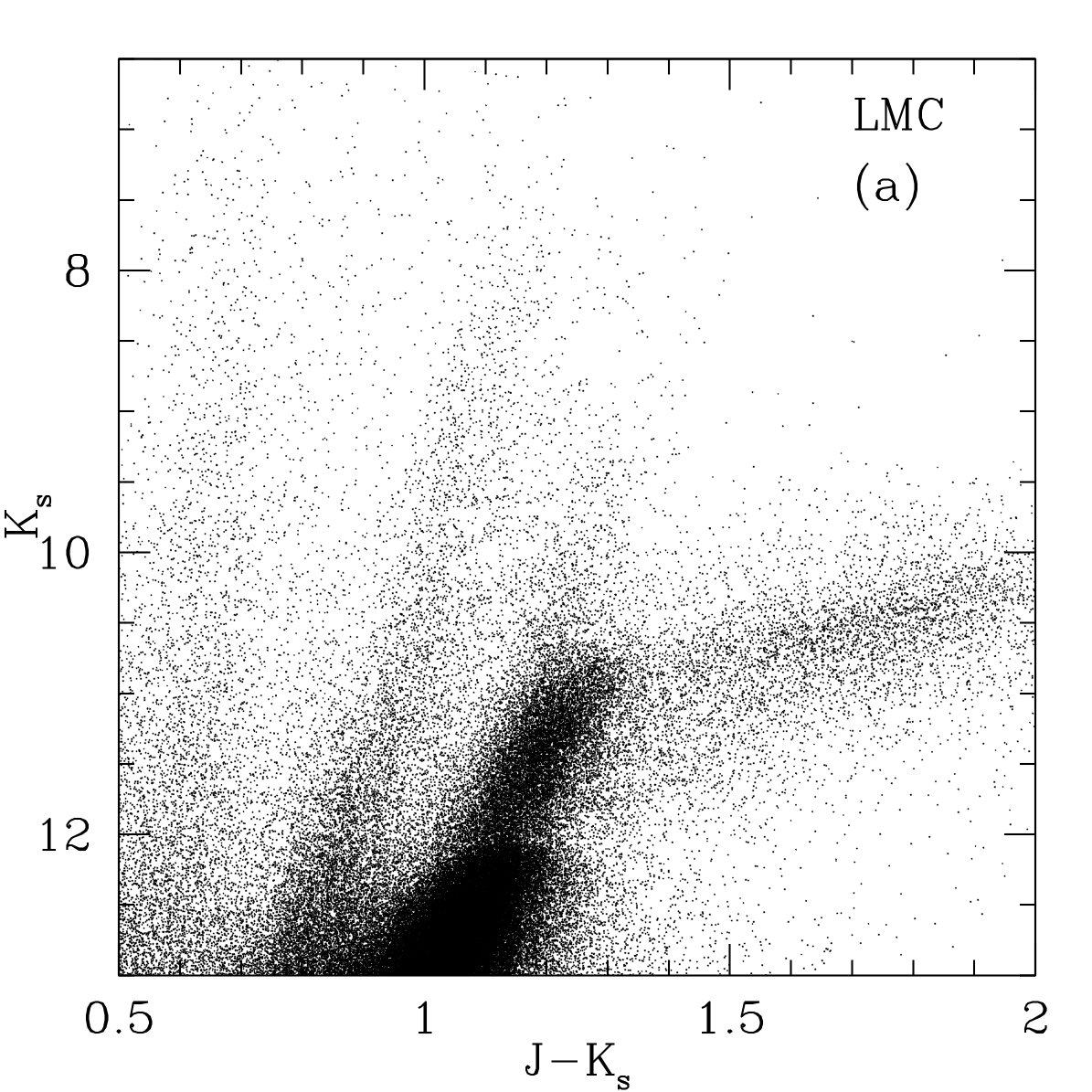}
\includegraphics[width=0.5\textwidth]{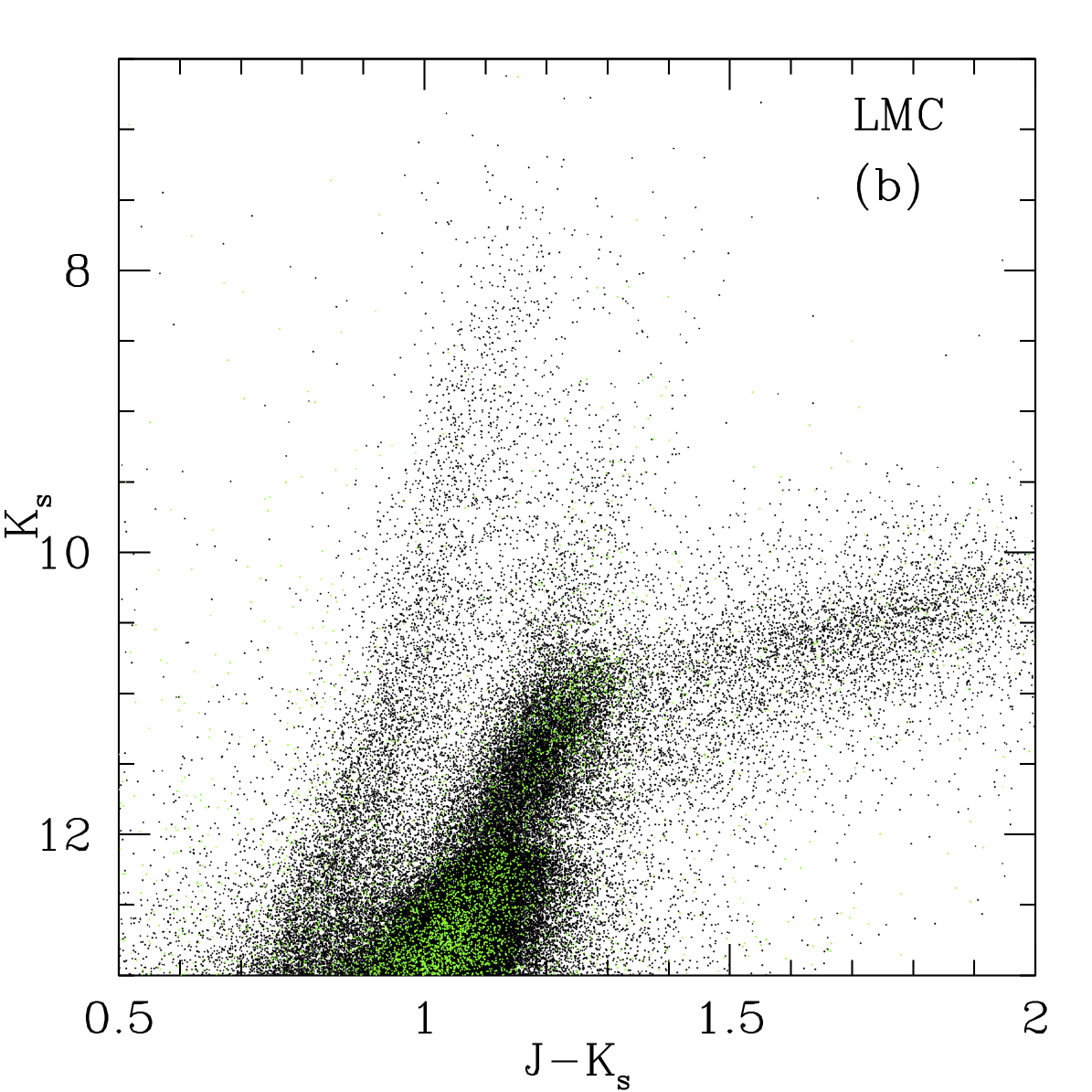}
\caption{\label{fig:CMD} The CMD for our sample. (a) The CMD is shown for all 87,637 stars in our initial sample obtain from 2MASS.  (b) The same as (a) but now with probable foreground stars removed.  The green points denote the stars either without any {\it Gaia} data or without {\it Gaia} parallax data.}
\end{figure}

In Figure~\ref{fig:CMD}(a) we show the CMD of the sample of 87,637 stars.  In the next two sections we will demonstrate how we refine these to select out only the RSGs.

\subsection{Removing Foreground Stars}
The majority of the very red stars in the LMC will be members, as previously shown by radial velocity studies (see, e.g., \citealt{NeugentLMC}), but a few will be foreground, and as we approach the warmer temperatures and yellower colors, there will be increasing contamination from Galactic stars. Indeed, in the color regime for YSGs, Galactic contamination becomes even more overwhelming. Fortunately, {\it Gaia} data \citep{Gaia} provides the means to identify these foreground objects through the judicious analysis of proper motions and parallaxes. We say ``judicious," as there is a known systematic offset in \emph{Gaia} DR2 astrometric measurements and the formal uncertainties provided do not represent the total error \citep{Lindegren2018}.

To assess the probability of LMC membership, we adopted a procedure similar to that described in \citet{Gaia}. First, we select a large set of highly-probable LMC stars in order to define the distributions of astrometric parameters that are expected for true members. We initially select all sources overlapping the LMC with \emph{Gaia} G $<$ 18 mag. In order to minimize any foreground contamination in this baseline sample, we then exclude all sources with either yellow colors (0.7 mag $<$ G$_{\mathrm{bp}}$ $-$ G$_{\mathrm{rp}}$ $<$ 1.1) or parallax measurements greater than 4$\sigma$ as well as any remaining sources whose parallaxes or proper motions deviated from the rest of the sample by $>$4$\sigma$. The red 2MASS stars identified above are then compared to this sample in order to assess their consistency with the kinematics of the  LMC. If the proper motions and parallax of a given star fall outside the region that contains 99.5\% of the comparison sample, we flag it as a probable foreground star. Conversely, if a star falls within the region that contains 75\% of the comparison sample, we consider it a likely LMC member. Stars that fall in between are flagged as having ``ambiguous'' membership based on astrometry alone. Further details of our application of this method can be found in \citet{ErinLBV}. Figure~\ref{fig:gaiaCuts} shows where our LMC members, probable foreground stars, and ambiguous results fall in proper motion and parallax space.

There were eight photometrically selected RSGs (as defined below) whose proper motions were consistent with LMC membership, but have quoted \emph{Gaia} DR2 parallax measurements that are more negative than our comparison sample of LMC stars. A large negative parallax is nonphysical, but suggests that these objects may not be foreground stars. We therefore retained these stars in our sample, but changed their status from ``probable foreground" to ``ambiguous" results. This included one star, 05300119-6956382, that had been previous identified in \citet{NeugentRSGII} as a RSG+B star binary.  

Two other RSG+B binaries from \citet{NeugentRSGII}, 05274747-6913205 and 05292143-6900202, would have also been dismissed as non-members were it not for the spectroscopic information.  In the case of 05274747-6913205, the Gaia parallax of 0.5561$\pm$0.0746 mas has a quoted significance of $>$7$\sigma$ and is consistent with a distance of 1.8 kpc (Bailer-Jones et al. 2018), but the proper motions and radial velocity (280 km s$^{-1}$) are in excellent agreement with membership in the LMC. The spectrum is that of a cool star with strong TiO bands, consistent with its J-K colors; Balmer lines are clearly present. We retain this star, but flag the {\it Gaia} results as ambiguous.  As for 05292143-6900202, \emph{Gaia} does not robustly detect a parallax (0.4757$\pm$0.1604); the proper motions are slightly outside the accepted spread we have adopted for membership, but the errors are large. Both its ground-based and {\it Gaia} radial velocities (also 280 km s$^{-1}$) suggest membership in the LMC. The spectrum is consistent with its colors, a late K or early M, with clear upper Balmer lines.  We also retain this star, describing its membership as ambiguous. We note that {\it Gaia} parallaxes can be impacted by both binarity and variability, both of which may be common in our sample.

One other star labeled as a RSG+B binary in \citet{NeugentRSGII}, 05065284-6841123, shows up as a foreground star. Further inspection of the unpublished AAT spectrum showed that there were reduction problems, and we no longer consider this star a RSG binary.

After cross-matching with {\it Gaia} (and making these small adjustments), out of the 87,637 red stars, 73,361 (83.7\%) were probable members; 3,585 (4.1\%) had ambiguous results; 9,651 (11.0\%) were probable foreground stars; and 1,040 (1.2\%) either had no match with {\it Gaia} or did not have {\it Gaia} parallax data that could be used to determine membership.

At this point, we removed the probable foreground stars from our sample but left the probable members as well as those with either ambiguous results or incomplete {\it Gaia} data. How the addition of the ambiguous results might alter our calculated binary fraction is discussed below. In Figure~\ref{fig:CMD}(b) we show the CMD after the foreground stars were removed. Note that the vast majority of these stars were those with lower $J-K_s$ values, consistent with our statement above that the contamination in our sample is primarily at the warmer temperatures. The green points are the stars for which there were incomplete or no {\it Gaia} data.
 
\subsection{Filtering Out AGBs and Red Giants}
Contamination by AGBs has long been the bane of RSG population studies. AGB stars are evolved low- to intermediate-mass stars which are in their He- and H-shell burning phase. These stars overlap in luminosity with RSGs below $\log L/L_\odot$ of 4.9 as noted by \citet{Brunish86}. Using optical photometry, one's only recourse was to limit RSG population studies to higher luminosities.

However, AGBs are cooler than RSGs since the Hayashi limit shifts cooler at lower masses \citep{HayashiHoshi}. \citet{Yang2019} used a ($J-K_s$, $K_s$) CMD to separate RSGs and AGBs in the SMC following the work of \citet{2006AA...448...77C} and \citet{2011AJ....142..103B}. \citet{UKIRT} adapted this method for their recent identification of RSGs in part of M31. They found that the color of the AGB/RSG boundary shifted in M31 relative to that of the SMC in the manner expected from the shifting of the Hayashi limit to cooler temperatures at higher metallicities demonstrating that cuts must be established for each galaxy separately.

\begin{figure}
\includegraphics[width=1\textwidth]{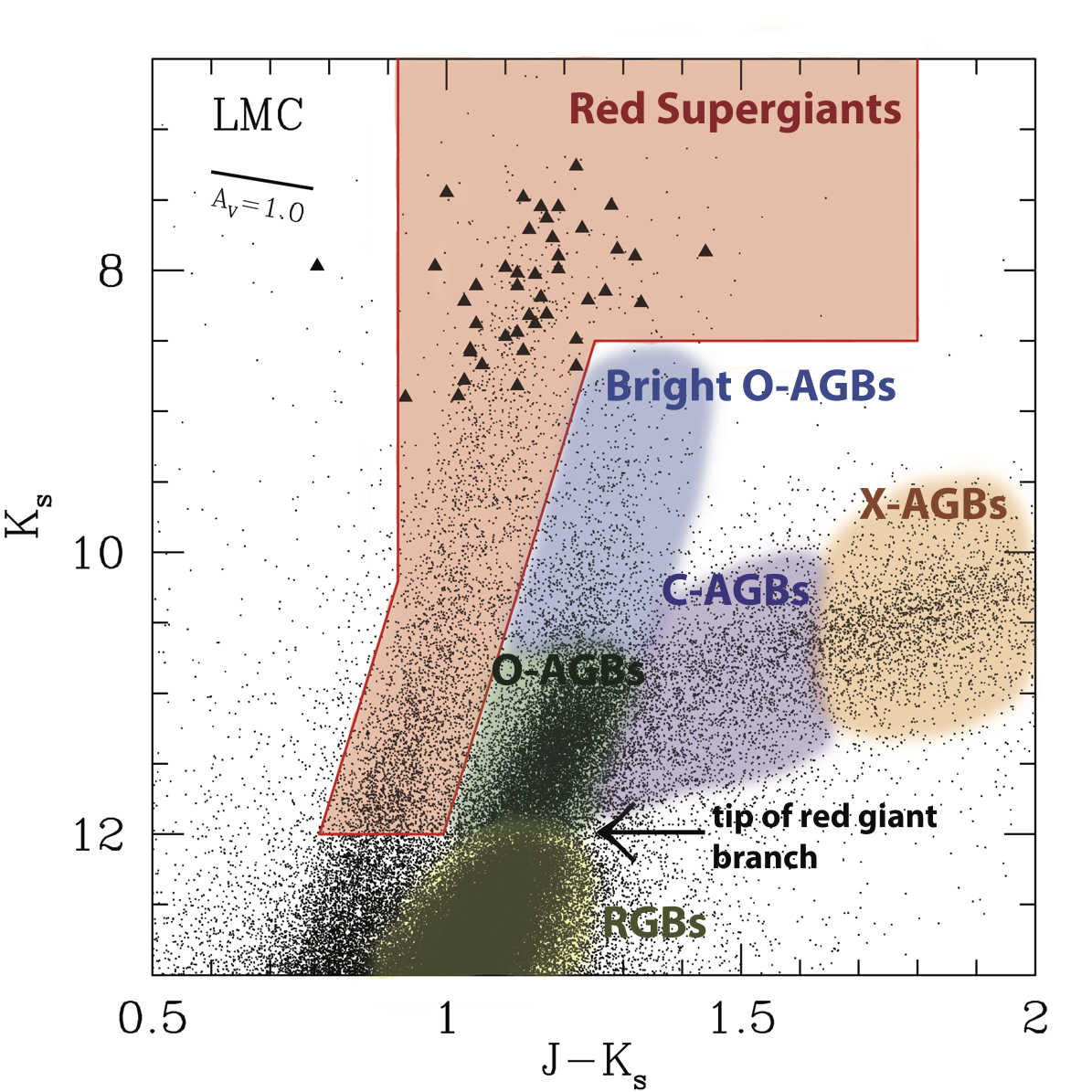}
\caption{\label{fig:wow} The CMD for cool members of the LMC. The various AGB branches \citep{2011AJ....142..103B} and red giant branch (RGBs) are labeled, along with the tip of the red giant branch.  (Note that the division between the carbon-rich AGBs [C-AGBS] and extreme AGBs [X-AGBs] is somewhat arbitrarily denoted in this diagram, as the actual definition was based upon $J-K$ colors \citealt{2011AJ....142..103B}.) The triangles show red supergiants analyzed from our previous work \citep{Levesque06,EmilyVariables,TZO}. The reddening vector corresponding to $A_V=1.0$ mag is also indicated.}
\end{figure}

\begin{figure}
\includegraphics[width=0.5\textwidth]{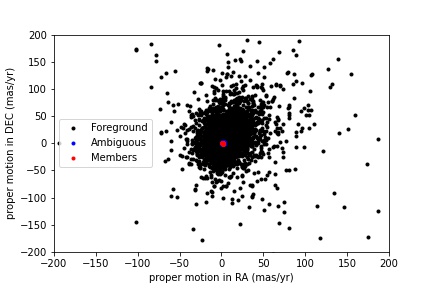}
\includegraphics[width=0.5\textwidth]{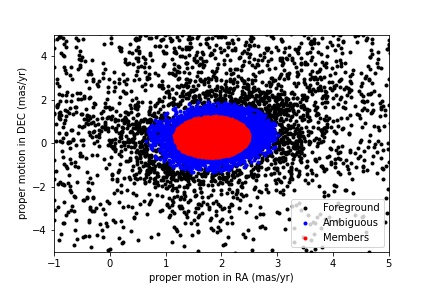}
\includegraphics[width=0.5\textwidth]{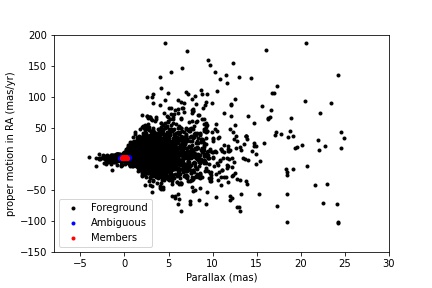}
\includegraphics[width=0.5\textwidth]{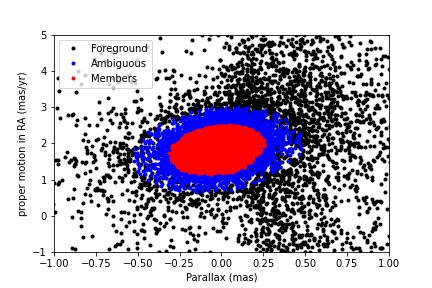}
\includegraphics[width=0.5\textwidth]{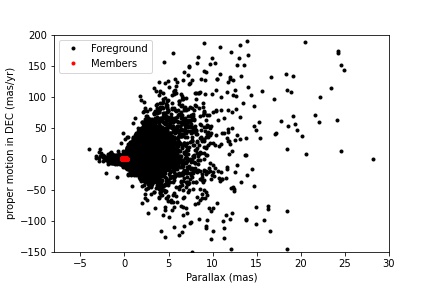}
\includegraphics[width=0.5\textwidth]{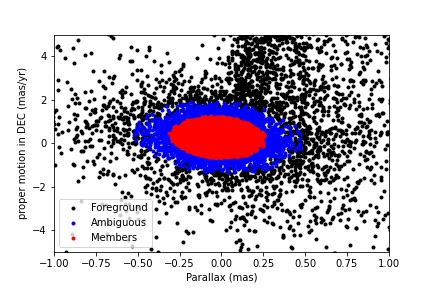}
\caption{\label{fig:gaiaCuts} {\it Gaia} information on LMC members, probable foreground and ambiguous stars. The top two figures show the proper motions in both right ascension (RA) and declination (DEC) plotted for all of the stars shown in Figure~\ref{fig:wow}'s CMD with the top right figure showing in a zoomed in version that better differentiates the differences between the three classification categories. The middle two figures show the proper motion in RA plotted against the parallax with again the bottom right figure showing a zoomed in version. The bottom two figures show the proper motion in DEC plotted against the parallax with again the bottom right figure showing a zoomed in version. Note that these figures were not used to {\it select} candidates, but rather simply show the results of our selections.}
\end{figure}

Here we repeat the same process for the LMC. Figure~\ref{fig:wow} shows the same CMD as shown previously, but now with the sequences labeled and probable foreground stars removed. The tip of the red giant branch (TRGB) is striking at $K_s=12$, or about $M_K=-6.5$\footnote{In contrast, the CMD shown by \citealt{UKIRT} for M31 (their Figure 8) goes to $K_s=17$, or $M_K\sim-7.6$, and so doesn't extend down as far as the TRGB.}. The location of the oxygen-rich, carbon-rich, and ``extreme" AGBs are shown, based upon the nomenclature of \citet{2011AJ....142..103B}; see, in particular, their Figure 4, based on a combination of 2MASS and Spitzer data. 

The realm of the RSGs is fairly easy to separate from these other stars, and we have drawn in the envelope of what we consider to be the RSG sequence. The locations are intermediate in color between what \citet{Yang2019} adopted for the SMC and what \citet{UKIRT} adopted for M31; this is consistent with the LMC having a metallicity that is intermediate between the two. (Note that the SMC and LMC metallicities are usually taken to be 1/3 and 1/2 solar, respectively; see \citealt{RussellDopita}, while M31's metallicity is about 1.5$\times$ solar; see \citealt{Sanders}.) In Table~\ref{tab:FunFacts} we provide the color relationships we use to define the RSG region of the LMC CMD. Next we'll describe how we defined our $K_s$ and $J-K$ cuts.

The red giant branch (RGB) and the RSG sequences begin to merge at $K_s=12.5$ and fainter which corresponds to $\log L/L_\odot \sim 3.3$. Since these diagrams go much fainter than our desired completeness limit of $\log L/L_\odot = 4$, we cut our RSGs at $K_s=12$, roughly the TRGB, similar to the approach adopted by \citet{Yang2019} for the RSGs in the SMC. This still allows completeness to $\log L/L_\odot \sim 3.6$ using the transformations in the next section, still considerably lower than the lowest luminosity we're concerned about.

After defining a faintness limit for $K_s$, we next investigated the best way of determining the lower value for $J-K$ as a function of $K_s$. When selecting RSGs in M31, \citet{UKIRT} chose to make their low (yellow) $J-K_s$ limit parallel to the high (red) $J-K_s$ limit following previous studies (e.g., \citealt{2011AJ....142..103B}). In their case, there were a substantial number of yellow stars without {\it Gaia} data and they were concerned with removing yellow foreground contamination. However, here we do not have this issue and placing similar cuts would impose unrealistic requirements on the temperatures of the brightest stars. Using the transformations in the next section, a $T_{\rm{eff}}$ of 4200~K corresponds to an observed $J-K_s=0.917$, where we assume a visual extinction $A_V=0.75$~mag as argued below. We therefore have modified the low (yellow) $J-K_s$ limit as shown in Figure~\ref{fig:wow} compared to what \citet{UKIRT} used in M31, and as documented in Table~\ref{tab:FunFacts}.  The scarcity of stars to the left of the low $J-K_s$ line is consistent with evolutionary theory: stars zip across the HRD to the RSG phase, spending very little time as YSGs, of order tens of thousands of years (see discussion in \citealt{DroutM31} and \citealt{NeugentLMC}). The tilt of this line at lower luminosities in essence says that higher luminosities RSGs are cooler than those of lower luminosities; this is consistent with what the evolutionary tracks say as well \citep{Ekstrom2012,BPASS2}.

Finally, we decided to relax the color requirement on the upper $J-K$ values at the brightest magnitudes, as there should no longer be any AGB contamination.  Again, this is consistent with what \citet{UKIRT} did in M31 and \citet{Yang2019} did for the SMC.  This allows for the fact that the higher luminosity RSGs could be more heavily reddened by circumstellar dust, an effect confirmed by \citet{UKIRT} in their M31 study. 

\subsection{Transformations}
To put our derived binary fraction in context, it is helpful to understand the physical properties of the stars probed; in addition, our LMC RSG sample will likely be used by ourselves and others for a variety of studies. We therefore provide here the transformations from the CMD to the physical HR diagram of $T_{\rm eff}$ and the log of the luminosity relative to that of the Sun ($\log L/L_\odot$). Our procedure closely parallels that of \citet{UKIRT}.

The typical OB star in the LMC has an extinction in the visual bandpass of $A_V=0.40$~mag \citep{LGGSII}, but in general, RSGs have larger extinction due to circumstellar dust \citep{Levesque05, Massey05}.  The typical $A_V$ for RSGs in the LMC is 0.75~mag based upon the spectrophotometric fits of 36 stars done by \citet{Levesque06}.  This is the same as we adopted for RSGs in M31 \citep{UKIRT} based upon the spectrophotometric fits of \citet{MasseySilva}.  In M31 we found a clear trend in $A_V$ with luminosity for the highest luminosity stars, not surprising given that those stars are likely to suffer higher mass loss (see, e.g., \citealt{Ekstrom2012}) as they approach the Eddington limit or develop late-stage pulsations at high luminosities \citep{vanLoon2005, Bonanos2010, Davies08}. Thus, we adopt $A_V = 0.75$ for the entire sample except for the few brighter stars ($K_s < 8.5$). For those stars, we determine the extra extinction in such a way to match the average $K_s$ vs $J-K_s$ relation along a reddening vector.  The relevant equations are given in Table~\ref{tab:FunFacts}.  These higher reddenings affected only 45 stars in our sample of 4090 RSGs (1.1\%), and had values that ranged from $A_V=1.34$~mag to 3.37~mag.  RSGs with considerably larger amounts of circumstellar extinction are known both in the Galaxy \citep{Massey05} and the LMC \citep{Levesque2009}.  The impact of these higher extinction values on the derived luminosities is relatively minor: as noted in Table~\ref{tab:FunFacts}, the equivalent $A_K$ values are only 12\% of the $A_V$ values, and at most the extra extinction we deduce increases  $\log L/L_\odot$ by 0.1~dex. Given the luminosity dependence on these higher mass loss events, we do not believe we are missing a population of lower luminosity RSGs with higher than expected reddening values.

Before applying any extinction correction, we first transform the 2MASS $J-K_s$ colors and $K_s$ brightness to the standard $J-K$ and $K$ system \citep{Bessell90} using the transformations determined by \citet{Carpenter}; these equations are given in Table~\ref{tab:FunFacts}. The transformation from $J-K$ to $T_{\rm eff}$ is then determined by first de-reddening the color assuming $E(J-K)=A_V/5.79$ \citep{Schlegel1998}, and then using the MARCS stellar atmosphere models \citep{MARCS} computed for LMC metallicity described by \citet{Levesque06} to relate the intrinsic $(J-K)_0$ colors to $T_{\rm{eff}}$. The typical errors on $T_{\rm{eff}}$ are 150~K, where this value is dominated not by the photometric uncertainties but rather by assuming an uncertainty of $\pm 0.5$ mag on our value for $A_V$. The relationship is quite linear over the relevant color range.  To determine the bolometric luminosity, we first correct the $K$-band photometry for extinction ($A_K=0.12A_V$, \citealt{Schlegel1998}).  The bolometric correction then comes from the adopted $T_{\rm{eff}}$, and we determine the bolometric magnitude using a distance modulus of 18.50 (50~kpc; \citealt{vandenbergh2000}).  The relevant equations are given in Table~\ref{tab:FunFacts}.

We note explicitly that our reddening correction makes the assumption of a normal \citet{CCM} reddening law with a ratio of total-to-selective extinction $R_V=3.1$.  However, we also note that our use of NIR photometry makes this assumption relatively benign, and our results robust. We have used $A_V$ to characterize the amount of reddening both for convenience and because the 0.75~mag value came out of fitting the optical spectrophotometry by \citet{Levesque06}. However, as shown in Table~\ref{tab:FunFacts}, we actually correct the photometry by $A_K$ (for luminosity) and by $E(J-K)$ for effective temperature and bolometric corrections to the luminosity. The relationships between these and $A_V$ come from \citet{Schlegel1998}, who adopted the \citet{CCM} law, which would be generally applicable in the optical and NIR to the interstellar dust found in the Milky Way and Magellanic Clouds. However, we know little about the dust properties of grains in the circumstellar environments of RSGs. As discussed by \citet{Massey05}, Galactic RSGs with abnormally large extinction compared to neighboring OB stars show a correspondingly large UV excess compared to stellar models, primarily indicative of scattering, but that large grains may also play a role. Indeed, the recent dimming of Betelgeuse seems like it was caused by a dust episode with grains that are so large that the extinction was nearly grey \citep{2020ApJ...891L..37L}. What we do know from multiple SED fittings, is that the CCM law works well at wavelengths beyond the near-UV; see, e.g., \citet{Levesque05, Levesque06, EmilyVariables}, even in cases of extremely high extinction, such as WOH G64 with $A_V=6.8$~mag \citet{Levesque2009}.  Thus, the assumption that the circumstellar reddening in the optical and NIR is similar to that of interstellar dust appears to be borne out empirically.  In addition, the use NIR photometry makes the issue of reddening relatively moot, given that the extinction in $A_K$ is only 12\% that of $A_V$, and thus an uncertainty even of 1~mag in $A_V$ would affect our $M_K$ value by only 0.12~mag. As mentioned above, such a mistake would affect the derived luminosity by 0.13~dex when the effect both on the extinction and bolometric correction were taken into account. It is indeed partially for this reason that we chose to use NIR photometry.

Table~\ref{tab:RSGContent} contains the coordinates, 2MASS $J$ and $K$ colors, and derived temperatures and luminosities
for the 4090 RSGs in the LMC. We note that the color limits imposed in the CMD require a $T_{\rm eff}$ as a function of $\log L/L_\odot$. For $4.0\leq \log L/L_\odot \leq 4.25$, $T_{\rm eff}$ has a minimum of $5300-362 \log L/L_\odot$ and a maximum of $5333-362\log L/L_\odot$. For $\log L/L_\odot>4.25$, $T_{\rm eff}$ has a minimum of 4200~K and a maximum of $5333-362\log L/L_\odot$.

\section{Spectroscopically Confirming RSG+B Binaries}
After selecting RSGs as described above, we next turned our attention towards identifying the subset of these RSGs that additionally have B star companions. To do this, we took a similar approach to the search for RSG binaries in M31 and M33 by \citet{NeugentRSGII} and used photometry to identify a subset of candidates before heading to Las Campanas for spectroscopic confirmation.

Overall, we obtained spectra of 63 candidates in the LMC. Of these, 25 were single RSGs and the remaining 38 were RSG+B binaries. We additionally observed 22 new candidates in the SMC and confirmed 14 as single RSGs and 8 as new RSG+B binaries which will briefly be discussed in \S3.3. 

\subsection{Selection Criteria}
While 2MASS NIR colors are helpful for identifying RSGs, we needed additional information about the RSGs' flux in the bluer wavelengths to determine if they have B star companions. For this information we used \citet{ZaritskyLMC}, which contains $U,B,V,$ and $I$ photometry for most of our survey region of 4090 stars. After cross-matching our list of LMC RSGs with \citet{ZaritskyLMC} using a 1\arcsec\ radius, we found that 3870 (95\%) had $U$-band photometry, 3992 (98\%) had both $B$ and $V$, and 3579 (88\%) had $I$. The remaining 98 stars had no match in \citet{ZaritskyLMC}, primarily due to crowding (over half of such stars were located in the inner bar).

Since RSGs with B star companions will have smaller $U-B$ colors (and thus higher flux at bluer wavelengths) than those without, we focused on observing RSGs with small $U-B$ colors. As discussed in \citet{NeugentRSGII}, we previously identified RSG+B star binaries in the Magellanic Clouds using archival spectra. Of the 23 we identified, 8 have $U-B < 0$, and all but 2 have $U-B < 1$. Thus, as we'll discuss when we calculate the binary fraction, we believe that the majority of RSG+B star binaries have $U-B$ colors less than 1 so we focused the majority of our spectroscopic observing efforts on those stars. However, we still observed a few candidates with $U-B$ colors between 1 and 2 to better characterize the binary fraction at slightly higher $U-B$ values and attempt to define a $U-B$ color ``cut-off" above which RSG+B star binaries aren't found.

Of the 95\% of LMC RSGs with $U$ and $B$ photometry, 127 (3\%) have $U-B < 0$, 388 (10\%) have $0 < U-B < 1$, and 2870 (74\%) have $1 < U-B < 2$. However, to maximize the number of stars observed, we focused on the brighter targets (generally those with $U$ brighter than 16th). This decreased our initial target list to 107 stars with $U-B < 0$, 142 with $0 < U-B < 1$, and 25 with $1 < U-B < 2$. We then attempted to observe a subset of those with a wide range of $U-B$ values to determine the binary fraction as a function of $U-B$.

\subsection{Observations and Reductions}
Candidate RSG+B star binaries were observed with the Magellan Echellette (MagE; \citealt{MagE}) instrument on the Baade 6.5-m telescope at Las Campanas Observatory over two dedicated observing runs in 2019 and 2020 and an engineering run in 2019. The first two night run occurred on UT 2019 September 07-08 when we observed both SMC and LMC targets and were plagued by atrocious seeing that varied between $2\farcs0$ and $4\farcs0$. We were still able to achieve adequate S/N by simply increasing our exposure times thanks to our objects' bright magnitudes ($U \sim 14.6, B \sim 14.3, V \sim 12.9$). The seeing was somewhat improved during our second run on UT 2020 January 14-15 with seeing that started out at $1\farcs0$ and degraded to $2\farcs2$. On the second run, just LMC targets were observed. Additionally, 14 LMC targets were observed with MagE during engineering time on UT 2019 September 12-13 with $1\farcs0$ seeing. On all runs we used a 1" slit and exposure times ranged between 300 for the brightest targets to 1200 for the dimmest targets obtained during poor seeing. The MagE instrument gives a wavelength coverage of 3400\AA\ to 1 $\mu$m at R $\sim4100$ allowing us to observe both the upper Balmer lines between $3700-4000$\AA\ and the TiO bands redwards of $6000$\AA\ simultaneously. We additionally observed spectrophotometric standards throughout the night to assist with flux calibration. The data were extracted using both the {\sc iraf} echelle package and {\it mtools} routines designed by Jack Baldwin for the reduction of spectra obtained with another one of Las Campanas' instruments, the Magellan Inamori Kyocera Echelle (MIKE).

\subsection{The Observed Sample}
Our goal when observing was both to spectroscopically confirm single and binary RSGs but also to get a sense of how the binary fraction might change with respect to increasing $U-B$ colors. To do this correctly, we had to be confident in our classifications and not, for example, mistakingly classify a single RSG as single when really we just hadn't observed long enough to detect the faint upper Balmer lines of its companion. Thus, our exposure times were dictated by our desire to either observe or conclusively rule out the presence of the upper Balmer lines. Based on previous observations described by \citet{NeugentRSG}, we determined that a S/N greater than 100 at our spectral resolution of $R \sim 4100$ was needed to definitively rule out the presence of upper Balmer lines coming from the faintest possible B star (a 15000 K B dwarf; $M_V = -1.5$). We therefore first observed each target with a short (5-10 minute) exposure and checked the S/N of the spectra in real time. We additionally performed quick-look reductions which were completed just a few minutes after each spectrum had read out. If the star showed upper Balmer lines, we moved on. If it didn't, and the S/N was below 100, we continued observing the candidate until we either detected the upper Balmer lines, or the S/N reached 100. Given this observing strategy, we are confident that the stars we have labeled as single do not have hidden B-star companions with $M_V > -1.5$, which should encompass all B-type stars.

Overall, the photometry of the spectroscopically confirmed binary and single stars was as expected with stars with lower $U-B$ colors being binaries. We observed 27 candidates with $U-B < 0$ and 74\% of them were RSG+B star binaries (the remaining 7 stars being single RSGs). We found a similar percentage of binaries (71\%) for the 24 stars we observed with $0 < U-B < 1$. For the remaining 12 stars we observed with $U-B > 1$, only one was a RSG+B binary. As discussed extensively in \S2.3.1 in \citet{NeugentRSGII}, likley reasons for single RSGs having anomalously blue $U-B$ colors include the possibility of dust scattering that produces a blue reflection nebula or even the much simpler explanation of poor initial photometry.

\subsection{Small Magellanic Cloud Observations}
While our overall goal was to determine the binary fraction of RSGs in the LMC, our first observing run was scheduled in early September and the LMC wasn't above 2 airmasses until around half way through the night. Thus, we started off each night by observing a few candidates in the SMC based upon stars with $U-B < 1$ and $U$ brighter than $\sim$ 16th from the RSG sample presented in \citet{Yang2019} and crossmatched with \citet{ZaritskySMC} for $U,B,V,$ and $I$ colors. Overall we observed 22 new candidates and confirmed 14 as single RSGs and 8 as new RSG+B binaries. Because we weren't able to observe a statistically significant sample of candidates, we are not comfortable estimating a binary fraction for RSGs in the SMC yet. While we hope to be able to expand on this research more in the future, at this point we've chosen to simply include our findings on these 22 stars as part of this paper in Table~\ref{tab:SMCtab}. A further discussion on deriving the physical properties of these stars can be found in the Appendix.

\section{Calculating the Binary Fraction}
To calculate the binary fraction of RSGs in the LMC, we followed a multi-step process. We first estimated an initial binary fraction using a K-nearest-neighbor algorithm (k-NN) that combined archival photometry with our spectroscopically observed single and binary RSGs. We then adjusted the fraction and corresponding error bars to account for the following biases: line-of-sight stars masquerading as binaries, binaries in eclipse not detected during the photometric survey, and RSGs in systems with non-B star companions. How we accounted for each of these biases is described below.

\subsection{Initial Estimate}
To produce the initial estimate of the binary fraction of RSGs with B-type companions based upon archival photometry and our spectroscopically observed LMC stars, we relied on a k-NN approach. This method is based upon the idea that stars with bluer colors and/or UV signal are more likely to be binary RSGs, which is something we've confirmed spectroscopically. The k-NN algorithm assigns probabilities of binarity to each of the remaining candidate stars that we did not spectroscopically observe by looking at how close they are in color-color space to the stars that have been spectroscopically confirmed. It follows that candidates with colors similar to known binaries are more likely to be binaries than those with colors similar to known single RSGs. This method allows us to calculate the percentage likelihood that each individual candidate is a binary RSG.

The input columns to the k-NN algorithm were all based on archival photometry including $U$,$B$,$V$, and $I$ photometry from \citet{ZaritskyLMC} as well as the calculated $U-B$ and $B-V$ values. We opted not to include the 2MASS $J$ and $K$ photometry because the single and binary RSGs were evenly distributed throughout the CMD and thus the NIR colors did not provide any additional information that could help classify the stars. We additionally included a flag related to the brightness of the star in the NUV based upon survey data from the GALaxy Evolution EXplorer ({\it GALEX}) \citep{Martin05, Morrissey07, Bianchi09}. RSG+B star binaries should be bright in the UV given the B star companion while single RSGs should not. Thus, we hoped that the presence of NUV signal would help identify binaries. To determine whether the {\it GALEX} data is sensitive to the lowest luminosity B companions, we determined whether we would detect the flux of a reddened LMC A0V with a typical magnitude of 21.3 in the {\it GALEX} NUV filter. According to \citet{Simons14}, the {\it GALEX} NUV detection limit in the LMC is 22.7 mag and thus we are sensitive to even the lowest mass B star companions.

NUV images for each of the 4090 LMC RSGs were downloaded from the Mikulski Archive for Space Telescopes (MAST) and then simple aperture photometry was run to obtain an estimate of the star's brightness. The stars were then grouped into four categories based on their aperture photometry: no data at the specified coordinates (24\%), data but no NUV signal detected (63\%), dim NUV signal (6\%), medium NUV signal (3\%), and bright NUV signal (2\%). The 76\% with aperture photometry were then visually checked to confirm that the category (none, dim, medium, or bright) matched what was found in the images. While the {\it GALEX} data proved to be very useful when used in combination with the \citet{ZaritskyLMC} photometry as part of the k-NN algorithm, it should be noted that there are issues present within the dataset. These are discussed in great detail in \citet{Simons14} but revolve around the {\it GALEX} resolution being quite large at 5" and thus inadequate in the crowded OB associations. Thus, we've used the {\it GALEX} data as one small piece of our overall method of determining binarity, and not as the determining factor. Still, we do find that confirmed RSG+B star binaries are brighter in {\it GALEX} than the confirmed RSG single stars with 68\% of the binaries with data showing either medium or bright NUV signal and 70\% of the single RSGs showing either dim or no NUV signal. Additionally, as discussed above, not all photometry (including \citealt{ZaritskyLMC}) is perfect. By using the k-NN approach and using inputs from different datasets in different passbands, we decrease the overall weight being placed on any individual measurement. Thus, we hope this will decrease erroneous results due to a single poor measurement of a star, for example.

To implement the k-NN algorithm, we relied on Python's \texttt{SCIKIT-LEARN} machine-learning package. Our total number of spectroscopically confirmed RSG+B binaries included the 36 described in this paper, as well as 10 discussed in \citet{NeugentRSGII}, 4 found by \citet{Levesque06} and 5 found by \citet{Dorda18} bringing the total up to 55. For single stars, we included the 23 described here, as well as 217 other spectroscopically confirmed single RSGs described in \citet{NeugentRSGII, Levesque06, Dorda18} and our own unpublished AAT data described in \citet{NeugentRSGII}. Thus, we had 295 stars we could use to both train and test our data. We first scaled our data using \texttt{SCIKIT-LEARN}'s RobustScaler to account for the fact that our features (magnitudes/colors, and flags) are in different units. We then used k-fold cross-validation to train and test our model. We found that splitting our data up into 8 folds (as opposed to the default 5) achieved the highest accuracy when re-run against the test dataset. Thus, each of the 8 test sets contained around 7 binaries and 27 single RSGs. During testing, we found that we had the highest success using a k-NN search that looked at the nearest 26 neighbors weighted based on distance. Using this method, we achieved an accuracy of 93.5\%. 

\begin{figure}
\epsscale{0.5}
\plotone{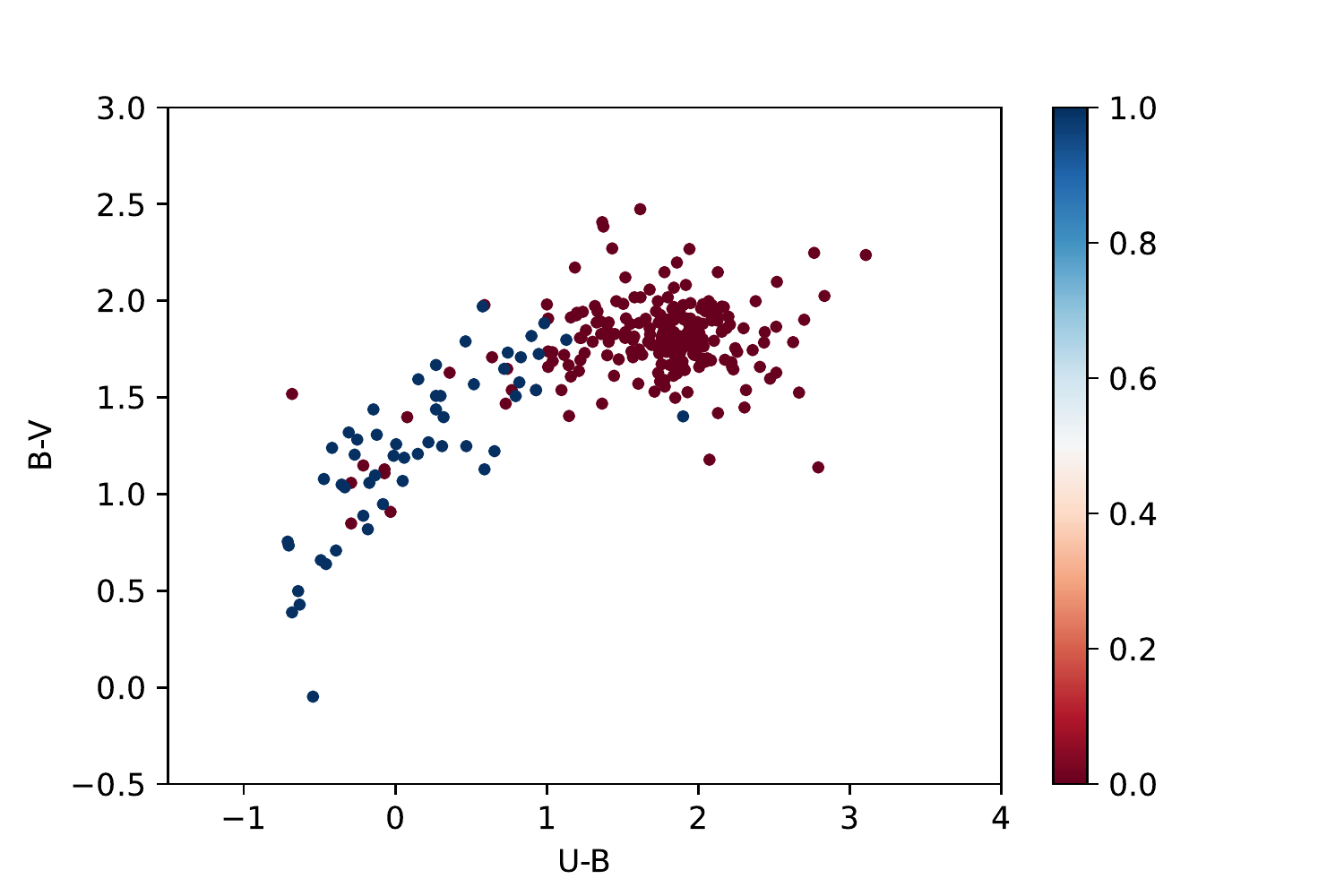}
\plotone{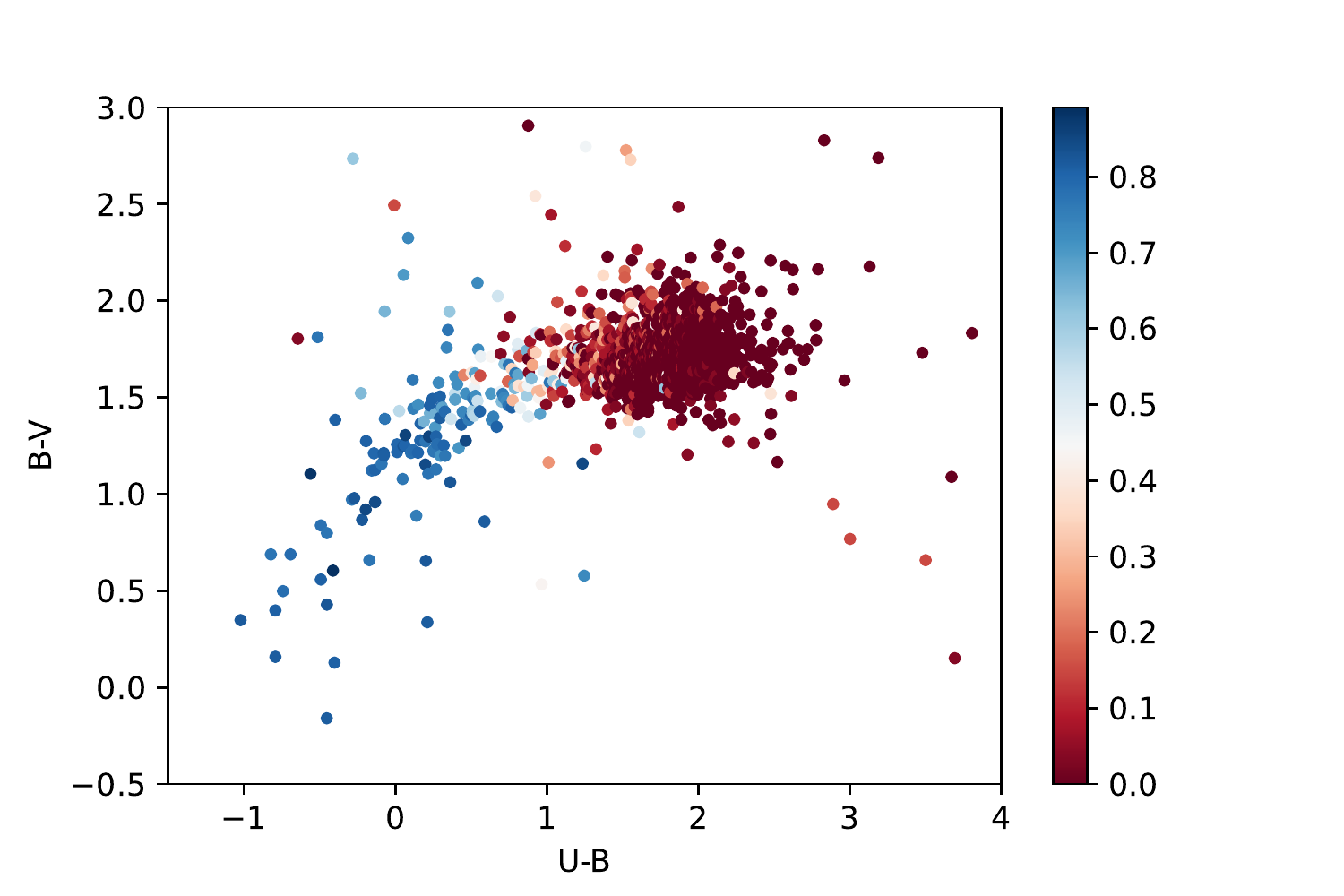}
\caption{\label{fig:KNN} Results from k-NN algorithm. The figure on the left shows the 295 spectroscopically confirmed single (red points) and binary (blue points) RSGs in color-color space. The figure on the right shows the results of the k-NN algorithm on the remaining candidate RSGs. The stars have been colored according to the percent likelihood of being a binary with the bluer points being more likely binaries and the redder points being more likely single RSGs.}
\end{figure}

We applied the k-NN algorithm to the 1457 stars in our sample with both $\log L/L_{\odot} > 4.0$ (our completeness limit) and a minimum of $B$ and $V$ photometry from \citet{ZaritskyLMC}. Figure~\ref{fig:KNN} shows a color-color plot of both the original input sample of 295 spectroscopically observed stars and the results from the k-NN classification run. It has been color-coded to reflect the percent likelihood of each star being a binary with the bluer points representing binaries and the redder points representing single stars. Note that the ``transitional-zone" is around a $U-B=1$, which is what we had concluded empirically during our observations. Stars with $U-B$ colors smaller than 1 are more likely to be RSG binaries while stars with $U-B$ colors higher than 1 are more likely to be single RSGs. The input data and final percentage likelihoods for each star are shown in Table~\ref{tab:kNN}.

Since we assigned a percent likelihood of binarity to each individual candidate star, we could then get a first estimate of the binary fraction, before taking any biases into account. By simply summing up the percent likelihood of each star being in a binary system and dividing by the percent likelihood of each star being a single RSG we can estimate the binary fraction of RSG+B stars with $\log L/L_\odot > 4$ ($M > \sim 9M\odot$). After adding in the 295 spectroscopically confirmed stars, and taking the 93\% accuracy rate of the k-NN algorithm into account, we arrive at a binary percentage of $13.5^{+7.56}_{-6.67}\%$ where the errors were calculated by assuming the most extreme scenarios given the 93\% accuracy rate. However, as we discuss next, there are other factors to take into account that will increase this percentage slightly.

\subsection{Eclipsing Binaries}
We additionally must consider eclipsing binaries since the majority of our binary fraction estimate is based upon single-epoch photometry. Take, for example, the Galactic RSG+B star binary system, VV Cep which has a 20.3 year orbit and is in secondary eclipse for 18 months, or around 7\% of the time \citep{Bauer2000}. If there are systems like VV Cep in our sample and the companion was behind the RSG when the photometry was obtained, these systems would not show up as binaries. Thus, we must account for this bias. If we had orbit determinations for our binaries, it would be possible to calculate this probability directly. However, since our classifications are based upon single-epoch spectroscopy, this is not possible. Instead we must make some assumptions about RSG binary systems and their orbits to determine what percentage of them are eclipsing at any given time. In the future, this calculation could hopefully be done independent of any modeling either by obtaining $U$,$B$, and $V$ photometry at another epoch (in essence, repeating the work of \citealt{ZaritskyLMC}), or a detailed analysis on the orbital parameters of our discovered RSG binaries. But at this point, this is outside the scope of the current work and thus can't be done observationally.

Instead, we turned to the BPASS models (v2.2.1) from \citet{BPASS1,BPASS2}. We are grateful to J.J. Eldridge for help providing a program that allowed us to easily estimate the percentage of RSGs that would be in eclipse at any given moment. BPASS uses the findings of \citet{Moe2017} as initial conditions and then evolves the binary systems to populate appropriate binary companions to RSGs and determines (along with a host of physical properties) their periods, mass ratios, and separations at an LMC-like metallicity of $z=0.008$. The maximum angle of inclination for eclipses is then computed based on the RSG's radius and the separation of the stars and then, for those that could possibly eclipse, the eclipse duration is determined. Simple Poisson errors were also estimated based on the number of eclipsing binaries. Based upon these calculations and the types of binaries we are sensitive to detecting, we estimate that $3.61\pm0.01\%$ of our targets that appear to be single RSGs are actually eclipsed binaries. (Note: Further information on how we ran BPASS to most closely align with our observations is discussed in \S5). This increases the binary fraction slightly, but not substantially.

\subsection{Line-of-Sight Pairings}
One possible contaminant in our survey is line-of-sight pairings. These are stars where both the RSG and B star are genuine LMC members, but the B star isn't gravitationally bound to the RSG and instead just happens to exist in our line of sight to the RSG. In general, we expect these cases to be rare given the photometric quality checks we went through when selecting the original list of RSG candidates in 2MASS (stars with poor photometry flags, and thus possible visual pairings, were ignored), but the presence of a faint B star in the LMC foreground or background could still contribute Balmer lines to a spectrum. (For example, take LGGS J004453.06+412601.7 which \citet{M31WRs} classified as a WN+TiO. We originally thought this might be the first example of a RSG+WR binary system, but based on archival imaging we determined it is actually a M31 WR + foreground M dwarf pairing). To determine the probability that each of our spectroscopically confirmed RSG+B star binaries are actually line-of-sight pairings, we ran a simple Monte Carlo simulation that took into account the OB star density around each of our confirmed RSG binaries. 

Overall, the process worked as follows: for each of our confirmed LMC RSG+B star binaries, we found the locations of the OB stars within a 5\arcmin\ radius of the binary using $B$ and $V$ photometry from \citet{ZaritskyLMC}. We then ran a simulation that randomly placed a RSG within this region and checked to see if it fell within 1\arcsec\ of one of the OB stars. If it did, we flagged it as a line-of-sight pairing. The value of 1\arcsec\ comes from the size of our slit while observing on Las Campanas. 

We opted to include both O stars as well as B stars in our simulation because, as described below, they are possible RSG binary companions, even if the likelihood is small. Also, given that we were relying on $B$ and $V$ photometry from \citet{ZaritskyLMC}, it is difficult to distinguish O and B stars from one another since their $B-V$ colors are nearly identical due to being on the tail end of the Rayleigh-Jeans distribution after having peak flux in the UV. To select the stars within the 5\arcmin\ radius of the binary, we removed everything redder than an A0V by using a cut at $(B-V) < 0.0$. We then took the average reddening of the LMC to be 0.13 from \citet{LGGSII} and set the brightness limit of $V = 21$ since spectroscopically we wouldn't be able to observe the upper Balmer lines from OB stars fainter than that.

After running the simulation 10,000 times, we found that there was a $1.9\pm2.0\%$ chance that any of the observed RSG+B star binaries was actually a line-of-sight pairing (with 0\% being the minimum and 10\% being the maximum for any individual system). We additionally ran the program on the spectroscopically confirmed single RSGs and found a very similar distribution with a $1.6\pm1.7\%$ chance that any of the single RSGs could have a line-of-sight companion (again 0\% was the minimum and 10\% was the maximum). Given both the similarity between these two results and their low values, we believe that line-of-sight pairings have a negligible impact on the overall binary fraction.

\subsection{RSGs+Other Companions}
In \citet{NeugentRSG,NeugentRSGII} we argue that RSGs will primarily have B-type companions from an evolutionary point of view because longer-lived main-sequence stars (A,G,K and M stars) won't have formed by the time a RSG is created. However, what about the shorter lived or non-main sequence stars such as O stars, YSGs, RSGs, WRs, etc.? From an evolutionary point of view, these systems are certainly possible. However, so far none have been observed and the lifetimes of such companion stars are so short that finding such a system is statistically unlikely. Here we delve deeper into each of these pairings and how their occurrence could alter our calculated RSG binary fraction.

The most likely system other than an RSG+B star binary is an RSG+O star binary, simply due to the longer duration of of an O star's time on the main sequence (a few million years) as compared to the later evolutionary stages (YSGs, RSGs, WRs, etc.) which last only tens to hundreds of thousands of years. Such a system would exist for a short period of time if two nearly equal mass stars were born together and one evolved into a RSG while the other was still on the main sequence. In terms of these stars biasing our calculated binary fraction, due to our photometric detection method, if any O star binaries do exist in the LMC, we've likely already detected them. The $U-B$ colors of O stars are nearly identical to that of B stars, so they would show up as candidates based on our k-NN algorithm (possibly even as higher likelihood candidates due to their lower $U-B$ values). We additionally would have been sensitive to them as spectroscopic candidates. Even though their upper Balmer lines have smaller equivalent widths, the stars themselves are brighter in $U$ and their strong He {\sc i} and He {\sc ii} lines would have shown up prominently. While none of our spectroscopically confirmed binaries appear to have O-type companions, our method of calculating the binary fraction is sensitive to such pairings.

Continuing on a massive star's evolutionary path are the even shorter-lived YSG and RSG stages. We admit that such systems would be difficult to detect and would likely only be observable if eclipsing or as a spectroscopic binary with some of the narrow metal lines (such as the Ca {\sc ii} triplet) appearing double. Again, since these pairings are statistically unlikely and have never been observed, we do not think they will effect our calculated binary fraction. However, we do point out that our method of detecting RSG binaries is not sensitive to such systems.

Next up are WR+RSG binary systems. We can confidently say that none of these have been detected in any of the nearby galaxies and furthermore, we don't expect to find any since the population of WRs in the LMC is thought to be complete \citep{FinalCensus}. Since the discovery method for finding the WRs was based on their strong emission lines, any WR+RSG binaries would have been found as part of these galaxy-wide searches. 

Finally, let’s consider RSGs with neutron star or, in the case of more massive primaries, black hole companions. Such systems can occur if the RSG is originally the less massive of the two stars and the binary is not disrupted when the primary explodes as a supernova (SN). If the post-SN separation is too small, the systems will subsequently interact, potentially merging to create a Thorne-\.Zytkow object such as the candidate recently found in the SMC \citep{TZO}. However, if the post-SN separation is wide enough, the secondary can expand into the RSG phase. While to the best of our knowledge no RSG+compact object binaries have yet been confirmed, and the majority of known high-mass X-ray binaries in the Magellanic Clouds have periods too short to allow the secondary to expand to the RSG phase \citep{Haberl2016,Antoniou2016}, there are a number of observational biases against detecting long-period systems. Indeed, RSGs have been identified as candidate counterparts/donor stars to several ultra-luminous X-ray sources \citep{Heida2016,Lopez2017} and recent high cadence time domain surveys are now facilitating the detection of non-interacting compact object systems \citep[e.g.][]{Thompson2019}. Theoretically, both the rate of binary disruptions and the post-SN separation distribution are highly dependent on uncertain SN kick prescriptions \citep[e.g.][]{Bray2018}. However, using BPASS v2.2.1 as described above, we estimate that $2.42\pm 0.01\%$ of RSGs have compact object companions.

\subsection{Final Binary Fraction}
We are now in the position to estimate the final binary fraction of RSGs in the LMC. We initially planned our observations to be sensitive to RSG+B star companions given that B-type stars should dominate the sample based on evolutionary constraints. Though, we additionally point out that our selection criteria is sensitive to the less common O-type companions as well. Using the k-NN approach described above, we observationally estimate the RSG+OB star binary fraction as $13.5^{+7.56}_{-6.67}\%$. We then used BPASS to estimate both the fraction of eclipsing binaries ($3.61\pm0.01\%$) and RSG+compact companions ($2.42 \pm0.01\%$) that we were not sensitive to in our search. Overall, we reach a final percentage of $19.5^{+7.6}_{-6.7}\%$ for RSGs with $\log L/L_\odot > 4$. These values are shown in Table~\ref{tab:fracs}. We stress that we are not including RSG+protostars in this calculation and that we are not sensitive to RSG+RSG or RSG+YSG systems, though from an evolutionary standpoint, these should be extremely rare. We expect from first principles that in order for a RSG+RSG system to exist even momentarily would require the two stars to be born with masses within 5\% of each other; for the RSG+RSG to last for the majority of the RSG phase of the higher mass star, it would require an initial mass ratio $q$ of 0.98-1.02.  A more exact calculation is beyond the scope of the present paper, but will be discussed in future work. In the next section we compare our results to expectations from BPASS for the total population as well as the binary fraction of other types of massive stars.

\section{Discussion}
Now that we've determined a binary fraction, we'd like to see where it fits within massive star observations and evolutionary theory. First we'll compare it to the binary fraction of other types of massive stars and discuss whether the number makes intuitive sense. Then we'll look at what the BPASS models predict, and finally we'll compare the physical properties of the single and binary RSGs before ending with a few words about our overall survey completeness. 

\subsection{Does this fraction match expectations?}
As discussed in the Introduction, the binary fraction of long-period, non-interacting OB-star systems could be between 70-100\% \citep{Gies08, SanaSci} with the short-period binary fraction being closer to 30-35\% \citep{Garmany80, Sana30Dor}. Since RSGs evolve primarily from OB stars, why is our calculated binary fraction of $19.5^{+7.6}_{-6.7}\%$ so much lower? 

The key thing to remember is that RSGs have radii that are hundreds to even thousands of times the radius of the Sun. Two main sequence stars less than $30M_{\odot}$ in a binary system must have separations on the order of thousands of solar radii to not interact at some point before the more massive star turns into a RSG, thus creating a RSG binary system. As is discussed in \citet{SanaSci}, binaries with orbital periods up to around 1500 days will exchange mass throughout their lifetime and, all except for one of the binary systems they measured had periods less than 1000 days (note: some of their rarity is a selection effect since long period systems are more difficult to detect spectroscopically but even in the ``corrected" sample set, there were few binaries with periods longer than 1000 days). Thus, the majority of these systems will interact before the more massive component turns into a RSG. When they interact, a few different things can occur. In close systems, RLOF will prevent the more massive star from ever turning into a RSG and a merger of the two OB-type stars might occur. In slightly more separate systems, the more massive star will turn into a RSG but then a merger between the evolved RSG and the unevolved companion might occur.

In short period systems, the two stars will begin interacting as the more massive star evolves and grows in radius. However, RLOF will eventually occur and the more massive star will be stripped of its entire envelope, losing much of its original mass. The secondary will then gain mass and angular momentum but neither will evolve into the RSG stage. As is discussed in \citet{SanaSci}, it is estimated that 40-50\% of O-star binaries will have their evolution altered due to RLOF.

In the case of a merger at the RSG phase, the binary system starts off with two main-sequence stars. Over time, the more massive star evolves first and eventually turns into a RSG with a companion. If the two stars are close enough, they will influence each other's orbits, begin spinning up, and transfer angular momentum. Once they merge, the RSG will photometrically appear single (though with a much higher rotational velocity). From photometry alone, a merged single star will generally be indistinguishable from an always-single star. While a RSG merger hasn't been directly observed, it has been hypothesized as an explanation for why one of the most famous RSGs is spinning so fast. Betelgeuse has a projected rotational velocity of around 15 km s$^{-1}$, much higher than that of a normal RSG. \citet{Wheeler17} suggest that this increased velocity could be due to a past merger with a smaller mass companion. \citet{SanaSci} estimate that 20-30\% of massive, apparently single stars, are actually the result of mergers.

Taking an initial O star binary fraction of 70\% and considering both mergers (20-30\% of binaries) and RLOF (40-50\% of binaries), our estimated binary fraction of $19.5^{+7.6}_{-6.7}\%$ is well in accord with the broad model predictions done by \citet{SanaSci}. Additionally, if we look at Fig.\ 1 (left) in \citet{SanaSci}, we see that $\sim15\%$ of the O star systems have periods longer than 1000 days, and thus would likely turn into RSG binary systems after the more massive star has evolved. This percentage is very well aligned with our findings.

\subsection{Comparison with BPASS Models}
As discussed in \S4, we used BPASS v2.2.1 \citep{BPASS1,BPASS2} to calculate the percentage of eclipsing binaries and RSG+compact companions. Here we go into a bit more detail about these BPASS simulations and compare the binary fraction we found to the BPASS results.

To ensure a fair comparison between our results and that of BPASS, we used our photometric selection criteria transformed to $T_{\rm eff}$ and $\log L/L_\odot$ (as described in \S2.4) to select model RSGs. We additionally placed a minimum mass constraint ($M > 8M_\odot$) on the RSG and minimum luminosity of $\log L/L_\odot > 4$. Using these two selection methods, we believe we've separated out the AGB stars in the BPASS models at least as well as we've done photometrically. Given these constraints, the BPASS models predict that 31\% are single RSGs, 25\% are merged RSGs, 2\% are RSGs + compact objects (as discussed above), and the remaining 42\% are RSG + main sequence star binaries (all percentages have Poisson errors $< 1\%$) at an LMC-like metallicity of $z = 0.008$. So, why the factor of 2 discrepancy?

The primary reason is that BPASS is a stellar evolution and population synthesis code and does not deal with star formation (yet!). Thus, all stars arrive on the zero age main sequence (ZAMS) at the same time, regardless of their mass. When taking the types of stars that might exist in binary systems with RSGs into account, BPASS uses the prescription given by \citet{Moe2017} without considering whether these stars would have arrived on the ZAMS by the time the RSG was formed. Due to the 
initial mass function (IMF) favoring lower-mass stars, this adds a significant number of low-mass companions.

\begin{figure}
\includegraphics[width=1\textwidth]{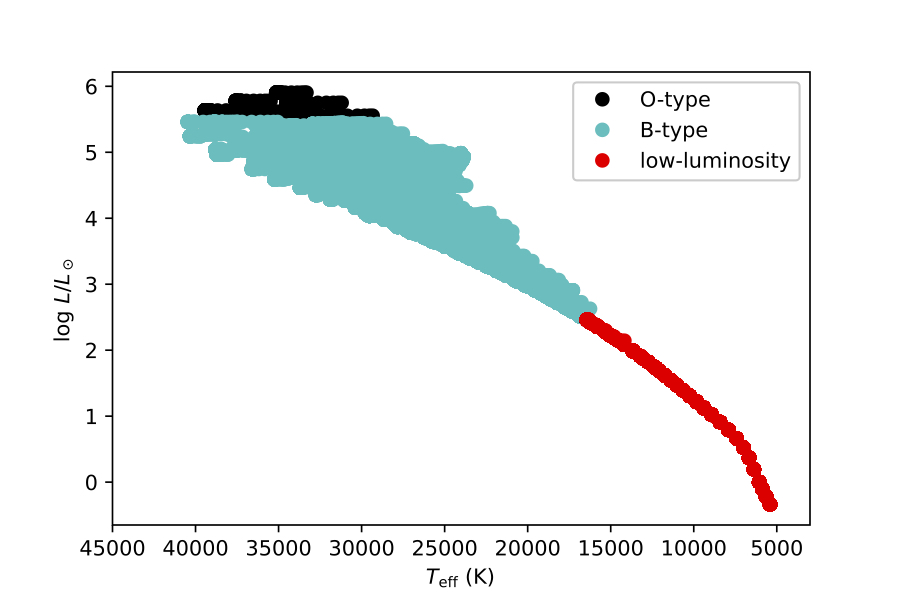}
\caption{\label{fig:BPASShrd} HR Diagram of RSG binary companions from BPASS v2.2.1. The O-type stars (black dots) make up less than 1\% of the sample, as is to be expected based on their short lifetimes while the B-type stars (cyan dots) make up the majority (74\%) of the sample. The lower luminosity stars (red dots) below a $\log L/L_\odot = 3.5$ will not have reached the ZAMS by the time the lowest mass RSG has formed and thus do not make viable companions.}
\end{figure}

Using the BPASS models, we can plot an HR diagram of the companions, as is shown in Figure~\ref{fig:BPASShrd}. From \citet{AAQ} we know that a B8V has a $\log L/L_\odot \sim 2.5$ and an B0I has a $\log L/L_\odot \sim 5.5$. Thus, we can make cuts in the BPASS companions to determine the percentage of binaries with O, B and less luminous companions. We find that, as expected based on lifetimes alone, O-type companions are exceedingly rare and make-up just 1\% of the binaries. The majority (74\%) are B-type stars, and the remaining 25\% are stars that wouldn't have reached the ZAMS before the formation of an RSG. Thus, we can conclude that 25\% of the RSG binaries estimated by BPASS are most likely single RSGs. This brings the BPASS-estimated binary fraction down to 32\% which includes RSGs with O and B-type companions, compact companions, and eclipsing systems. Since our calculated fraction is lower, this may suggest (again) that either the merger fraction is underestimated or the initial OB binary fraction is overestimated. 

\subsection{Physical Properties of Single vs.\ Binary RSGs}
\begin{figure}
\epsscale{1}
\plotone{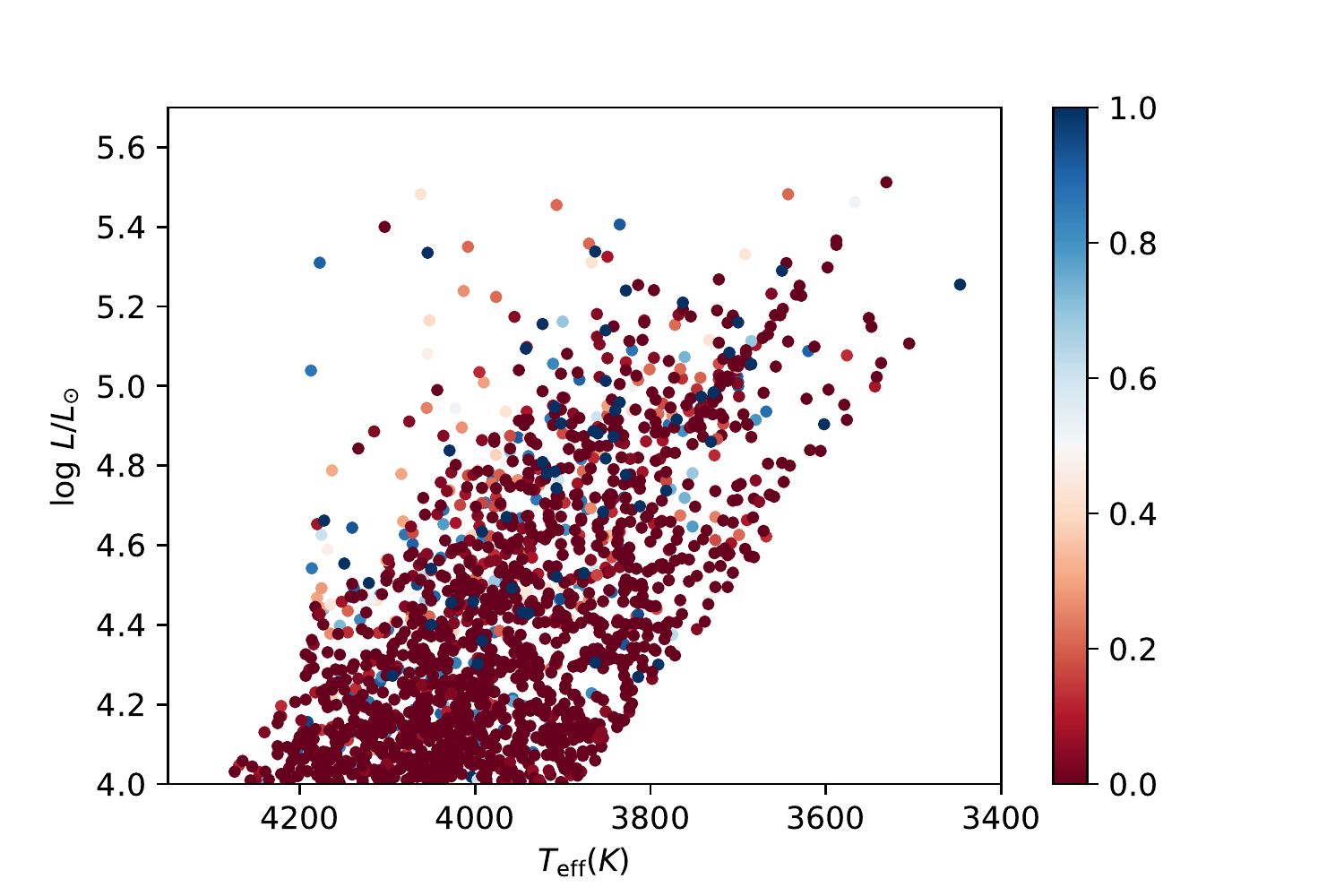}
\caption{\label{fig:HRD} HR Diagram of our RSG sample based on photometrically determined $T_{\rm eff}$ and luminosities. As in Figure~\ref{fig:KNN}, the stars have been colored according to the percent likelihood of being a binary with the bluer points more likely binaries and the redder points more likely single RSGs.}
\end{figure}

As described in the Appendix, we additionally obtained estimates of $T_{\rm{eff}}$, luminosities, and radii for the 63 LMC RSGs we observed on Magellan. We can additionally estimate the radii of our entire sample of k-NN classified RSGs using the photometrically calculated temperatures and luminosities, as shown in Figure~\ref{fig:HRD}. Since stars with larger radii will interact with a larger fraction of binary companions, it is interesting to examine both the average radii of single vs. binary stars and the binary fraction of large vs. small radius stars. Since these parameters will be impacted by a variety of factors, from the initial binary fraction as a function of mass to the IMF, we compare these parameters found for our sample to the same parameters from the BPASS models.

First, we can compare the physical properties of the 25 single RSGs to those of the 38 binaries we observed spectroscopically. One might expect that the average radii of the binaries will be smaller since RSGs with larger radii are more likely to have merged with their companions. Since the temperatures of RSGs are relatively constant because they sit at the Hayashi limit, it follows that the average luminosities of RSGs in binary systems might be lower as well. However, simply averaging the temperatures, luminosities and radiis for both the single and binary systems shows that there is no difference between the two sets. The average $T_{\rm{eff}}$, $\log L/L_{\odot}$, and $R$ for the spectroscopically confirmed binaries is $3710\pm80$K, $4.75\pm0.25$, and $610\pm220R_{\odot}$, respectively. For spectroscopically confirmed single stars, it is $3700\pm100$K, $4.78\pm0.25$, and $640\pm210R_{\odot}$.

We can also divide the k-NN classified sample into two bins, each with around 730 stars: those with small RSG radius ($R < 300R_\odot$) and those with large RSG radius ($R > 300R_\odot$). A simple calculation reveals that, surprisingly, the binary fraction is much higher (36\% vs.\ 4\%) for RSGs with higher radius. Overall, we'd expect that larger radii RSGs should have a higher incidence of mergers, but this might also be balanced out by the steep dependence on the IMF. If we do a similar study with the BPASS RSG+OB binary systems, we find that there is almost no dependence on RSG radius and in both bins (small and large RSG radius), the binary fraction is around 30\%. As we continue to spectroscopically observe more single and binary systems, we will find out whether our observed results are due to small number statistics, or whether the two populations really differ that drastically.

\subsection{Completeness Issues}
There were six additional stars we observed spectroscopically that were not included in the binary fraction calculations for various reasons. Each of these reasons points to a possible completeness issue (field size, 2MASS flag requirements, and color-cuts), which we discuss in detail below. However, we believe that each of these issues will simply lower our sample's overall completeness of RSGs in the LMC, but should in all cases impact binaries and single stars in the same manner, thus having negligible impact on our final binary fraction.

As described in \S2.1, we chose to select RSGs within a well-defined region of the LMC centered on $\alpha_{\rm J2000}$=05:18:00 and $\delta_{\rm J2000}$=-68:45:00 and extending in radius by 210\arcmin. This region was chosen based on our previous survey of WRs in the LMC and because it covers the entire optical disk of the galaxy. However, while we believe this region encompasses the majority of RSGs within the LMC, there are certainly a few outside of this region. Two such examples are 2MASS stars 04415417-6727202, a confirmed RSG+B star binary, and 05535411-6647126 a single RSG. As discussed recently by \citet{Nidever19}, the size of the LMC is an ongoing topic with fainter stellar streams being found continuously. While our radius selection means that we've missed some of the RSGs on the outskirts, there is no physical reason why we would be missing single stars or binaries preferentially and thus, this doesn't alter the determined binary fraction.

When selecting candidates using 2MASS, we additionally only kept those with the best photometry (quality flags of ``AAA," and ``artifact contamination" flags of ``000"). However, based on lists included in \citet{Dorda18}, we observed two candidates with lower quality flags. 2MASS star 05254453-6616228 had a flag of EAA and turned out to be a RSG+B binary and 05402532-6915302 had a ``ddd" flag and is a single RSG. As with our size selection, our method of choosing flags will hinder our completeness since we will miss a few RSGs with sub-standard quality flags, but binaries and single stars will be equally incomplete and thus this will not change the binary fraction.

Finally, there are two stars that fell outside our color cuts in $K$ and $J-K$. These cuts are always going to be difficult to execute perfectly due to uncertain star-by-star reddening values and thus it is not unexpected that there will be a few RSGs a bit redder than our cut or even one or two a bit bluer, such as the early K-type stars. Two examples are the single RSG 05312818-6703228 which was slightly too blue with a $J-K$ of 0.916 instead of the required 0.917 and the binary RSG 05401638-6659303 which was a little too red. To verify that our $J-K$ cuts weren't altering our overall binary fraction, we measured the fraction as a function of $J-K$ and found it to be constant to within our errors.

We also wanted to assess whether the ambiguous {\it Gaia} data might have changed our completeness rate or the binary fraction. There were initially 3,585 stars in our sample with ambiguous results (4.1\%). After filtering out AGBs, and making the appropriate color cuts, only 4 stars remained in our sample with $\log L/L_\odot > 4$. Two of them were spectroscopically confirmed LMC RSGs and the remaining two were classified as single RSGs by the k-NN algorithm. The classification of these two stars, even if they turn out to be foreground red dwarfs, will not change our final binary fraction.

\newpage
\section{Summary and Next Steps}
Here we observationally constrained the RSG binary fraction in the LMC to $19.5^{+7.6}_{-6.7}\%$ for stars with $\log L/L_\odot > 4$ corresponding to RSGs $> 9M\odot$. We did this by first identifying a complete sample of LMC RSGs using 2MASS NIR color-cuts and filtering out foreground stars using {\it Gaia}. In total we identified 4090 RSGs with $\log L/L_\odot > 3.5$ and 1820 with $\log L/L_\odot > 4.0$, which we believe to be our completeness limit. We then observed a sample of these spectroscopically to confirm their single vs.\ binary status. Since the binaries will have excess flux in the blue coming from the B star component, we then used photometry to determine binarity. Combining $U,B,V$ and $I$ photometry from \citet{ZaritskyLMC} and NUV brightness from {\it GALEX}, we used a k-NN approach to estimate the binary fraction of RSGs using our spectroscopic sample as a training set. From this approach we calculated a base binary fraction of RSG+OB stars as $13.5^{+7.56}_{-6.67}\%$. Our observations were not sensitive to either binaries in eclipse or RSGs in systems with compact companions so we used BPASS to calculate these percentages as $3.61\pm0.01\%$, and $2.42 \pm0.01\%$, respectively. Overall, we reach a final percentage of $19.5^{+7.6}_{-6.7}\%$ for RSGs with $\log L/L_\odot > 4$. This percentage does not include RSGs in systems with protostars or the rare case of RSGs in systems with other RSGs or YSGs. We then compared our result to what was discussed in \citet{SanaSci} and BPASS v2.2.1 modeling results. Our results are consistent with the broad expectations based on \citet{SanaSci} binary fractions, but slightly lower than the 32\% predictions by detailed calculations of BPASS.

In the future, we hope to decrease the errors on our observational measurements by spectroscopically confirming more RSGs as either single or binaries and thus increasing the accuracy of our kNN algorithm by building up our training set. This can be done by focusing on obtaining spectra of stars in the ``transitional" zone in $U-B$ space between 0 and 2 where it is not entirely clear whether a star should be labeled single or binary. Observations are planned using GMOS on Gemini-S for Fall 2020 to accomplish this goal.

We have similar spectroscopic and photometric data for the galaxies of M31, M33 and the SMC and next plan on determining the binary fraction of RSGs in those environments. Given their different metallicities, we additionally hope to determine whether there is a metallicity dependence on the binary fraction of RSGs. Finally, we are also observationally investigating the merger fraction of RSGs. Overall, we hope to determine the fraction of single RSGs, merged RSGs, and binary RSGs across a wide range of metallicities.

\acknowledgements
The authors first thank J.J. Eldridge for help interpreting the BPASS models as well as thoughts on some of the RSG+less luminous binary systems. We are also grateful to Konstantina Boutsia for enabling us to observe some of our targets during MagE Baade engineering and for obtaining herself two of the spectra. We also thank Trevor Dorn-Wallenstein for suggesting the k-NN machine learning algorithm to estimate the RSG binary fraction and Mario Juric for providing comments on how to determine the line of sight binary fraction. We additionally thank the support staff at Las Campanas for their always-excellent assistance as well as the anonymous referee for suggestions that improved the paper. The authors acknowledge that much of the work presented was done on the traditional land of the first people of Seattle, the Duwamish People past and present and honor with gratitude the land itself and the Duwamish Tribe. This work was supported in part by NSF IGERT grant DGE-1258485 as well as by a Cottrell Scholar Award from the Research Corporation for Scientific Advancement granted to EML. P.M.’s work was supported by the National Science Foundation under grant AST-1612874. MRD acknowledges support from the NSERC through grant RGPIN-2019-06186, the Canada Research Chairs Program, the Canadian Institute for Advanced Research (CIFAR), and the Dunlap Institute at the University of Toronto.

This work made use of the following facilities and software:

\facility{Las Campanas Magellan Telescopes, Galaxy Evolution Explorer, 2MASS, {\it Gaia}}

\software{IDL, IRAF (distributed by the National Optical Astronomy Observatory, which is operated by the Association of Universities for Research in Astronomy under a cooperative agreement with the National Science Foundation), Matplotlib v3.1.1 \citep{Hunter:2007}, NumPy v1.17.2 \citep{numpy:2011}, Pandas v0.25.1 \citep{pandas:2010},  Python 3.7.4, Scipy v1.3.1 \citep{scipy:2001}}

\appendix
Although secondary to the current project, we used the newly collected spectra to determine the physical properties of our sample.  \citet{Levesque05,Levesque06} describes the method of fitting {\sc marcs} models synthetic spectra to the observed optical spectra, using the depths of the TiO bands to as the primary temperature indicator\footnote{The use of the TiO band strength as an effective temperature indicator  has been challenged by \citet{Davies2013}, who argue that SED fitting is preferable.  However, SED fitting is sensitive to the adopted reddening law, and it is well established that circumstellar dust introduces significant complications \citep{Massey05}.  Furthermore, broad-band photometric SED temperatures rely upon an exact reproduction of the effective bandpasses, which is not straightforward (see \citealt{Bessell98}). Finally, the strengths of the TiO bands do, after all, form the basis of the spectral classification of RSGs \citep{1973ARA&A..11...29M}, and the resulting revision in the $T_{\rm{eff}}$ scale brought about by the Levesque et al.\ studies resulted in excellent agreement between the location of RSGs in the HRD and those predicted by evolutionary theory (see, e.g., \citealt{Ekstrom2012}).  Most importantly, the temperatures derived from TiO band strengths track the shifting of the Hayashi limit (cooler than stars no longer in hydrostatic equilibrium,  \citealt{HayashiHoshi}) to warmer temperatures with decreasing metallicity \citep{Levesque06,MasseySilva}. This is a fundamental expectation of stellar astrophysics; see discussion in \citet{LevesqueRSGs}.}.  We list these $T_{\rm{eff}}$ values in Table~\ref{tab:temps}.

\begin{figure}
\epsscale{0.5}
\plotone{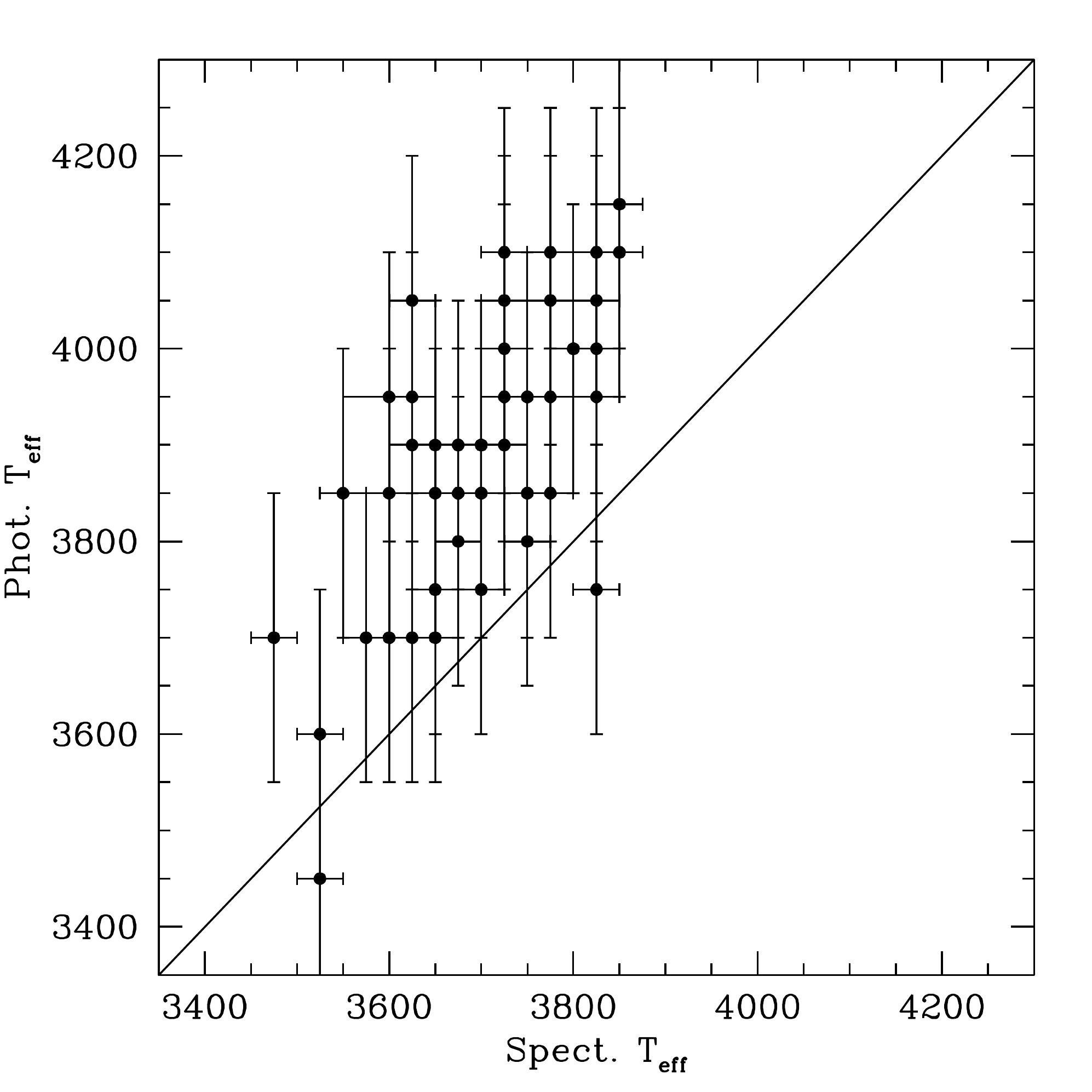}
\caption{\label{fig:Tcomp} Comparison of temperature determinations.  The photometrically determined $T_{\rm{eff}}$ values are plotted against the spectroscopically determined temperatures.  As found by \citet{Levesque05} there is a systematic issue, with the {\sc marcs} models giving higher (200~K) temperatures based upon the SED particularly for the warmer stars.  The line shows the one-to-one relation.}
\end{figure}

As shown by \citet{Levesque05} and \citet{MasseySilva},  the {\sc marcs} models are inconsistent in the sense that the $T_{\rm{eff}}$ values derived from $V-K$ photometry are systematically higher than those derived from fitting the spectophotometry by $\sim$150~K, particularly for the warmer (earlier-type) RSGs.  The vast majority of the stars in our sample have temperatures derived from $J-K$ photometry, and we thought it would be useful to show a comparison.  For ease, we include our photometrically determined temperatures in Table~\ref{tab:temps} as well. We show the comparison in Figure~\ref{fig:Tcomp}. Although the two agree within the 1$\sigma$ errors, there is again a systematic offset, with the {\sc marcs} models giving a higher temperature than those based upon the SED, particularly for the warmer stars.  We note that the photometric errors are dominated by the uncertainty in the adopted extinction, which we assumed was $A_v=0.75\pm0.5$ as discussed earlier. 

The fitting process also determines the $A_V$ directly for the single RSGs; the composite optical SED is too badly affected
by the blue color of the companion to be able to make an accurate determination of $A_V$ for the binaries.  For the photometrically determined temperatures, we simply adopted $A_V=0.75$, based upon the LMC stars fit by \citet{Levesque05}.  How do the spectroscopically determined $A_V$ compare with this value?  The average $A_V$ from the spectroscopy of single RSGs is 0.68, with a standard deviation of the mean of 0.08.  (The median value is 0.62.)  The 0.07~mag difference between the spectroscopically determined $A_V$ and the adopted one is negligible in terms of the physical parameters we derive, {\it on average}, translating to a difference in $(J-K)_0$ of only 0.01~mag.

\begin{figure}
\epsscale{0.5}
\plotone{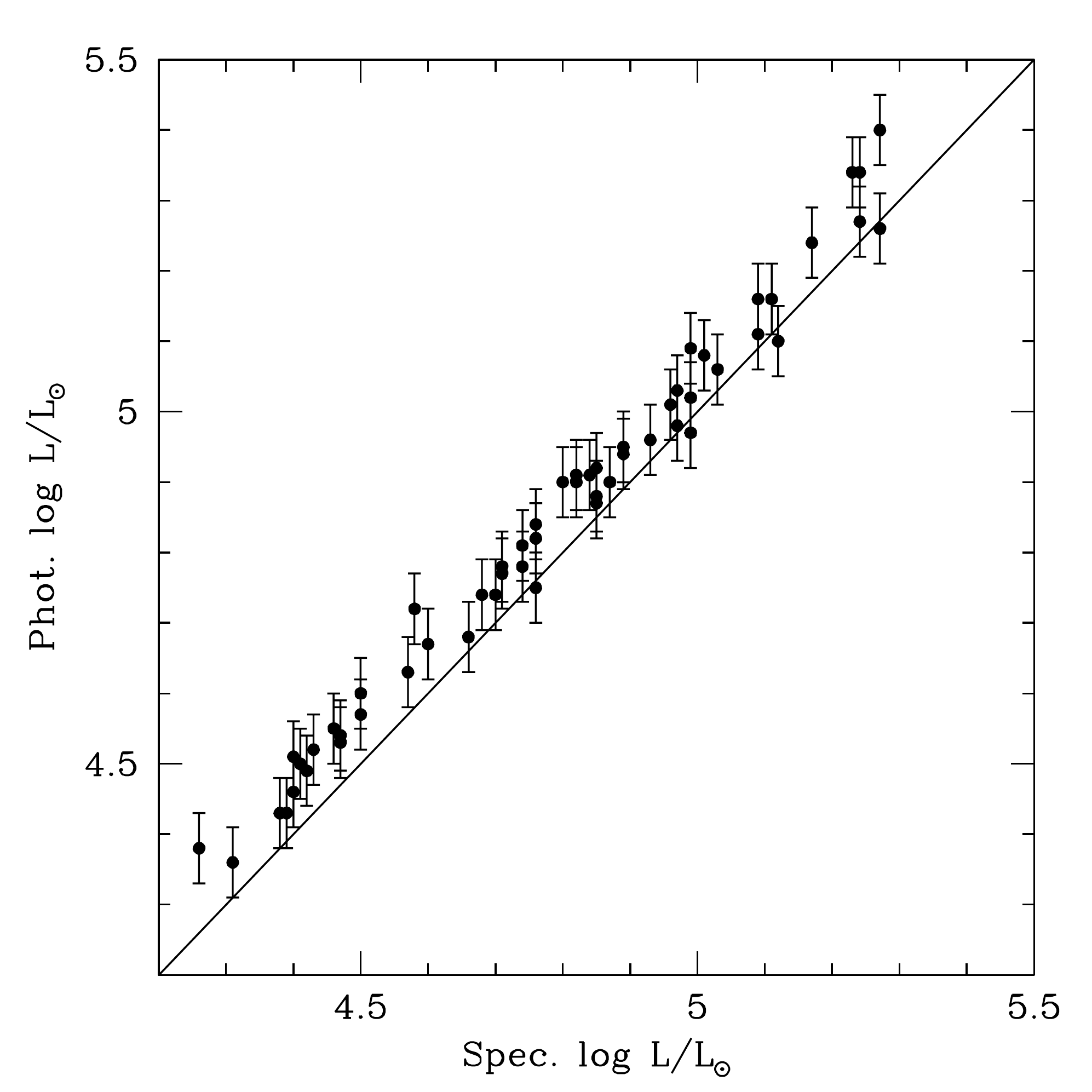}
\caption{\label{fig:Lcomp} Comparison of RSG luminosities.  The photometrically determined luminosities are plotted against the spectroscopically determined luminosities.  Despite the differences in $T_{\rm{eff}}$ and different determinations of the extinction, there is little difference in the two. The line shows the one-to-one relation.}
\end{figure}

What effect does the difference in methodologies have on the luminosities, which are, after all, what we are primarily interested in?  We show the comparison in Figure~\ref{fig:Lcomp}. Despite the offset in $T_{\rm{eff}}$, there is very little difference in the bolometric luminosities.  The differences are comparable to the 0.05~dex uncertainties in the luminosities determined photometrically.

\newpage
\bibliographystyle{apj}
\bibliography{masterbib}

\newpage
\begin{deluxetable}{l l r}
\tabletypesize{\scriptsize}
\tablecaption{\label{tab:FunFacts} Adopted and Derived Relations}
\tablewidth{0pt}
\tablehead{
&\colhead{Relation}
&\colhead{Source}
}
\startdata
\sidehead{Adopted Distance:}
\phantom{MakeSomeSpace} & LMC: 50~kpc & 1 \\
\sidehead{Reddening Relations:}
&$A_K =0.12 A_V = 0.686 E(J-K)$ & 2 \\
&$E(J-K) = A_V/5.79 $ & 2 \\
\sidehead{RSG Photometric Criteria:}
        &$10.20<K_s \leq 12.0$: $K_s \geq K_{s0}$ and $K_s \leq K_{s1}$ & 3 \\
        &$K_s\leq10.20$: $J-K_s\geq 0.917$ and $K_s \leq K_{s1}$ & 3 \\
        &$K_s\leq 8.5$ and $(J-K_s)\leq 1.8$: $J-K_s\geq 0.917$ & 3 \\
        &$K_{s0}=22.62-13.542(J-K_s)$  & 3,4\\
        &$K_{s1}=25.46-13.542(J-K_s)$  & 3,4\\
\sidehead{Adopted Extinction:}
        & $K_s>8.5$: $A_V=0.75$ & 3 \\
        & $K_s\leq 8.5$ and $K_s\leq K_1$: $A_V=0.75$ & 3 \\
        & $K_s\leq 8.5$ and $K_s\geq K_1$: $A_V=0.75+5.79\times\Delta(J-K_s)$ & 3 \\
        & $\Delta(J-K_s)=(J-K_s) - (24.04-K_s+0.686 (J-K_s))/14.228$ & 3 \\
\sidehead{Conversion of 2MASS ($J, K_s$) to Standard System ($J, K$):}
&$K=K_s + 0.044$ & 5\\
&$J-K = (J-K_s+0.011)/0.972$ & 5\\
\sidehead{Conversion to Physical Properties (Valid for 3500-4500 K):}
&$T_{\rm eff} = 5606.6 - 1713.3 (J-K)_0$ & 3\\
&${\rm BC}_K = 5.495 - 0.73697 \times T_{\rm eff}/1000$ & 3\\
&$K_0 = K - A_K$ & \nodata \\
&$M_{\rm bol} = K_0 + {\rm BC}_K - 18.50$ & 1 \\
&$\log L/L_\odot =(M_{\rm bol}-4.75)/-2.5$ & \nodata \\
\enddata
\tablerefs{1--\citealt{vandenbergh2000}; 2--\citealt{Schlegel1998}; 
3--This paper; 4--\citealt{2006AA...448...77C}; 5--\citealt{Carpenter}
}
\end{deluxetable}

\begin{deluxetable}{l l l r r r r c c r c c}
\tabletypesize{\scriptsize}
\tablecaption{\label{tab:RSGContent} Red Supergiant Content of the LMC}
\tablewidth{0pt}
\tablehead{
\colhead{2MASS}
&\colhead{$\alpha_{2000}$}
&\colhead{$\delta_{2000}$}
&\colhead{$K_s$}
&\colhead{$\sigma_{Ks}$}
&\colhead{$J-K_s$}
&\colhead{$\sigma_{J-Ks}$}
&\colhead{Gaia\tablenotemark{\scriptsize{a}}}
&\colhead{Spect.\tablenotemark{\scriptsize{b}}}
&\colhead{$A_V$}
&\colhead{$T_{\rm eff} [K]$\tablenotemark{\scriptsize{c}}}
&\colhead{$\log L/L_\odot$\tablenotemark{\scriptsize{d}}}
}
\startdata
04393719-6856276&04 39 37.194&-68 56 27.63&10.816&0.019&1.033&0.031&0&0&0.75&4000&3.97 \\
04394815-6935580&04 39 48.158&-69 35 58.01&11.825&0.024&0.937&0.033&0&0&0.75&4150&3.62 \\
04395031-6846522&04 39 50.313&-68 46 52.28&11.619&0.021&0.903&0.032&0&0&0.75&4200&3.72 \\
04395844-6849535&04 39 58.440&-68 49 53.58&11.994&0.021&0.903&0.032&3&0&0.75&4200&3.57 \\
04400185-6916490&04 40 01.854&-69 16 49.04&10.977&0.023&0.949&0.035&0&0&0.75&4150&3.95 \\
04401895-6941085&04 40 18.952&-69 41 08.57&11.857&0.023&0.999&0.035&0&0&0.75&4050&3.57 \\
04402177-6835339&04 40 21.771&-68 35 33.95&11.379&0.023&0.973&0.032&0&0&0.75&4100&3.78 \\
04404852-6822211&04 40 48.526&-68 22 21.15&11.242&0.023&0.876&0.035&0&0&0.75&4250&3.88 \\
04405219-6804580&04 40 52.194&-68 04 58.00&11.014&0.019&1.064&0.030&0&0&0.75&3950&3.87 \\
04410088-6840425&04 41 00.880&-68 40 42.53&11.833&0.023&0.821&0.033&2&0&0.75&4350&3.67 \\
\enddata
\tablenotetext{*}{This table is published in its entirety in the machine-readable format. A portion is shown here for guidance regarding its form and content.}
\tablenotetext{a}{LMC membership based upon {\it Gaia:} 0=member, 1=uncertain, 2=incomplete or no data, 3 = ambiguous.}
\tablenotetext{b}{Spectroscopy used: 0=no spectra, 1=CTIO 4-m from \citet{Levesque05}, 2=Magellan data from \citet{NeugentRSGII}, 3=Magellan data (this paper).}
\tablenotetext{c}{Typical uncertainty 150~K.}
\tablenotetext{d}{Typical uncertainty 0.05 dex.}
\end{deluxetable}

\begin{deluxetable}{l l l r r r r r r r l r r l}
\tabletypesize{\scriptsize}
\tablecaption{\label{tab:SMCtab} Spectroscopically Observed SMC Stars}
\tablewidth{0pt}
\tablehead{
\colhead{2MASS}
&\colhead{$\alpha_{2000}$}
&\colhead{$\delta_{2000}$}
&\colhead{$K_s$}
&\colhead{$\sigma_{Ks}$}
&\colhead{$J-K_s$}
&\colhead{$\sigma_{J-Ks}$}
&\colhead{$U$}
&\colhead{$B$}
&\colhead{$V$}
&\colhead{Class.}
&\multicolumn{3}{c}{RSG Component}\\ \cline{12-14}
& & & & & & & & & & &
\colhead{$T_{\rm eff} [K]$}
&\colhead{$\sigma_{T_{\rm eff}}$}
&\colhead{Type}
}
\startdata
00473688-7304441 & 00 47 36.886 & -73 04 44.18 & 8.319 & 0.024 & 1.147 & 0.033 & 15.603 & 14.734 & 12.736 & RSG & 3525 & 25 & M2\\
00503842-7319359 & 00 50 38.420 & -73 19 35.95 & 10.206 & 0.025 & 0.963 & 0.032 & 16.018 & 15.547 & 14.054 & RSG+B & 3825 & 100 & K5-M0\\
00523496-7226017 & 00 52 34.968 & -72 26 01.73 & 10.023 & 0.023 & 0.855 & 0.033 & 15.073 & 14.574 & 13.256 & RSG & 3875 & 100 & K5-M0\\
00523564-7251053 & 00 52 35.650 & -72 51 05.32 & 9.745 & 0.023 & 0.854 & 0.031 & 15.553 & 14.594 & 13.096 & RSG & 3875 & 100 & K5-M0\\
00531772-7246072 & 00 53 17.729 & -72 46 07.20 & 9.271 & 0.023 & 1.005 & 0.033 & 14.103 & 13.884 & 12.836 & RSG & 3800 & 100 & K5-M0\\
00532528-7215376 & 00 53 25.290 & -72 15 37.68 & 9.758 & 0.023 & 0.942 & 0.033 & 15.293 & 14.744 & 13.296 & RSG+B & 3850 & 100 & K5-M0\\
00534156-7215268 & 00 53 41.563 & -72 15 26.83 & 9.590 & 0.023 & 0.954 & 0.033 & 14.953 & 14.624 & 13.226 & RSG & 3900 & 100 & K2-3\\
00534451-7233192 & 00 53 44.517 & -72 33 19.21 & 9.462 & 0.020 & 1.021 & 0.030 & 14.563 & 14.464 & 13.226 & RSG+Be & 3725 & 25 & K5-M0\\
00562532-7228182 & 00 56 25.324 & -72 28 18.26 & 10.032 & 0.021 & 0.888 & 0.030 & 14.823 & 14.634 & 13.376 & RSG & 3950 & 100 & K2-3\\
00585831-7213429 & 00 58 58.310 & -72 13 42.93 & 9.837 & 0.021 & 0.923 & 0.032 & 13.463 & 13.894 & 13.066 & RSG & 3900 & 100 & K2-3\\
00595187-7243351 & 00 59 51.870 & -72 43 35.15 & 9.528 & 0.019 & 0.981 & 0.029 & 15.703 & 14.824 & 13.236 & RSG+B & 3875 & 100 & K5-M0\\
01004445-7159389 & 01 00 44.454 & -71 59 38.96 & 9.911 & 0.026 & 0.963 & 0.039 & 15.545 & 14.944 & 13.531 & RSG+B & 4050 & 100 & K2-3\\
01012693-7201414 & 01 01 26.930 & -72 01 41.43 & 9.235 & 0.024 & 0.991 & 0.034 & 14.863 & 14.434 & 12.926 & RSG & 3900 & 100 & K2-3\\
01014357-7238252 & 01 01 43.579 & -72 38 25.29 & 9.358 & 0.021 & 1.005 & 0.031 & 14.323 & 14.364 & 13.076 & RSG+B & 3900 & 100 & K2-3\\
01020407-7226109 & 01 02 04.076 & -72 26 10.90 & 9.420 & 0.023 & 1.035 & 0.033 & 14.623 & 14.614 & 13.286 & RSG & 3775 & 100 & K5-M0\\
01024480-7201517 & 01 02 44.801 & -72 01 51.75 & 9.386 & 0.021 & 0.954 & 0.030 & 15.613 & 14.634 & 12.986 & RSG & 3900 & 100 & K5-M0\\
01033730-7158448 & 01 03 37.301 & -71 58 44.88 & 9.598 & 0.020 & 0.962 & 0.030 & 15.013 & 14.484 & 13.096 & RSG+B & 3825 & 100 & K5-M0\\
01033984-7239059 & 01 03 39.849 & -72 39 05.93 & 10.362 & 0.021 & 0.969 & 0.032 & 16.065 & 15.436 & 13.955 & RSG+B & 3850 & 100 & K5-M0\\
01034536-7207490 & 01 03 45.360 & -72 07 49.03 & 9.639 & 0.025 & 0.959 & 0.034 & 14.963 & 14.764 & 13.386 & RSG & 3850 & 100 & K5-M0\\
01061197-7214380 & 01 06 11.970 & -72 14 38.00 & 10.072 & 0.019 & 0.901 & 0.029 & 14.633 & 14.484 & 13.346 & RSG & 4000 & 100 & K2-3\\
01064766-7216118 & 01 06 47.669 & -72 16 11.85 & 8.312 & 0.019 & 0.929 & 0.031 &  &  & 11.870 & RSG & 3750 & 25 & K5-M0\\
01081478-7246411 & 01 08 14.787 & -72 46 41.10 & 9.174 & 0.023 & 0.955 & 0.033 & 15.263 & 14.314 & 12.696 & RSG & 3850 & 100 & K5-M0\\
\enddata
\tablenotetext{*}{$J$ and $K$ photometry from 2MASS. $U,B,V$ photometry from \citealt{ZaritskySMC}.}
\end{deluxetable}

\begin{deluxetable}{l l l l l r r r r r}
\tabletypesize{\tiny}
\tablecaption{\label{tab:kNN} Percent Likelihood of Binarity\tablenotemark{\tiny{*}}}
\tablewidth{0pt}
\tablehead{
\colhead{2MASS}
&\colhead{$U$\tablenotemark{\tiny{a}}}
&\colhead{$B$}
&\colhead{$V$}
&\colhead{$I$}
&\colhead{$U-B$}
&\colhead{$B-V$}
&\colhead{NUV Flag\tablenotemark{\tiny{b}}}
&\colhead{Spec. Flag\tablenotemark{\tiny{c}}}
&\colhead{Binary \%}
}
\startdata
04411972-6935466 & 18.17 & 16.33 & 14.55 & 12.87 & 1.85 & 1.78 & 0 & 0 & 0\\
04411983-7006575 & 17.71 & 16.08 & 14.48 & 12.84 & 1.63 & 1.60 & 0 & 0 & 0\\
04412336-6851303 & 17.88 & 16.09 & 14.45 & 12.77 & 1.80 & 1.63 & 0 & 0 & 0\\
04423661-6817567 & 16.87 & 15.61 & 13.77 & 11.49 & 1.26 & 1.84 & 3 & 0 & 14\\
04425441-6826500 & 16.27 & 15.82 & 14.21 & 12.31 & 0.45 & 1.62 & 4 & 0 & 23\\
04431303-6947187 & 18.37 & 16.54 & 14.93 & 13.03 & 1.84 & 1.60 & 4 & 0 & 3\\
04432439-6855342 & 17.64 & 15.54 & 13.78 & 11.84 & 2.09 & 1.76 & 4 & 0 & 0\\
04433893-6946464 & 18.09 & 16.23 & 14.50 & 12.89 & 1.86 & 1.73 & 0 & 0 & 0\\
04434250-6758042 & 17.64 & 16.19 & 14.15 & 11.54 & 1.46 & 2.03 & 4 & 0 & 0\\
04434290-6746555 & 17.35 & 15.68 & 13.82 & 11.92 & 1.67 & 1.86 & 4 & 0 & 0\\
04434579-6932204 & 17.55 & 15.37 & 13.74 & 11.88 & 2.18 & 1.62 & 4 & 0 & 0\\
04441164-6906054 & 17.46 & 15.22 & 13.45 & 11.67 & 2.24 & 1.77 & 0 & 0 & 0\\
04441474-6948013 & 17.95 & 15.98 & 14.23 & 12.57 & 1.96 & 1.75 & 4 & 0 & 0\\
04443117-7012430 & 17.77 & 15.43 & 13.64 & 11.93 & 2.34 & 1.80 & 0 & 0 & 0\\
04443612-7043022 & 16.24 & 15.36 & 13.80 & \nodata & 0.88 & 1.56 & 4 & 0 & 46\\
\enddata
\tablenotetext{*}{This table is published in its entirety in the machine-readable format. A portion is shown here for guidance regarding its form and content.}
\tablenotetext{a}{$U$,$B$,$V$, and $I$ Photometry from \citealt{ZaritskyLMC}.}
\tablenotetext{b}{{\it GALEX} NUV brightness: 0 = n/a, 1 = bright flux, 2 = medium flux, 3 = dim flux, 4 = no flux.}
\tablenotetext{c}{Spectra Flag: 0 = no spectra, 1 = spectra.}
\end{deluxetable}

\clearpage
\begin{deluxetable}{l c c r r l r r r r r r r r l}
\tabletypesize{\tiny}
\tablecaption{\label{tab:temps} Comparison of Physical Properties}
\tablewidth{0pt}
\tablehead{
\colhead{2MASS}
&\colhead{$\alpha_{2000}$}
&\colhead{$\delta_{2000}$}
&\colhead{$V$\tablenotemark{\tiny{a}}}
&\colhead{$K_s$\tablenotemark{\tiny{b}}}
&\colhead{Class}
&\multicolumn{3}{c}{Photometry}
&
&\multicolumn{5}{c}{Spectroscopy} \\ \cline{7-9} \cline{11-15}
&&&&&&
\colhead{$A_V$}
&\colhead{$T_{\rm eff}$ [K]\tablenotemark{\tiny{c}}}
&\colhead{$\log L/L_\odot$}
&
&\colhead{$A_V$\tablenotemark{\tiny{d}}}
&\colhead{$T_{\rm eff}$ [K]\tablenotemark{\tiny{e}}}
&\colhead{$\log L/L_\odot$}
&\colhead{$R/R_\odot$}
&\colhead{Sp.Type}
}
\startdata
04415417-6727202&04 41 54.170&-67 27 20.20& 13.46&  8.07&RSG+B & \nodata& \nodata& \nodata&& 0.75&3525&  4.93& 780&M3      \\
04490536-6747133&04 49 05.360&-67 47 13.30& 12.04&  7.80&RSG+B &    0.75&    3900&    5.16&& 0.75&3725&  5.09& 850&M0      \\
04501563-6835019&04 50 15.631&-68 35 01.98& 13.58&  9.33&RSG+B &    0.75&    3900&    4.53&& 0.75&3700&  4.47& 420&M1      \\
04523565-7040427&04 52 35.659&-70 40 42.72& 13.01&  8.59&RSG+B &    0.75&    3850&    4.82&& 0.75&3650&  4.76& 600&M1.5    \\
04524274-6922061&04 52 42.743&-69 22 06.18& 13.54&  9.60&RSG+B &    0.75&    4000&    4.46&& 0.75&3800&  4.40& 360&K5-M0   \\
04543854-6911170&04 54 38.547&-69 11 17.00& 13.25&  7.20&RSG+B &    0.75&    3450&    5.26&& 0.75&3525&  5.27&1160&M3      \\
04551604-6919120&04 55 16.049&-69 19 12.08& 12.88&  7.37&RSG   &    0.75&    3700&    5.27&& 0.75&3625&  5.24&1050&M2      \\
04561441-6623167&04 56 14.419&-66 23 16.72& 13.50&  9.49&RSG+B &    0.75&    3950&    4.49&& 0.75&3725&  4.42& 390&M0      \\
04561739-6627297&04 56 17.392&-66 27 29.70& 12.83&  8.34&RSG   &    0.75&    3850&    4.92&& 0.04&3675&  4.85& 660&M1      \\
04562363-6942110&04 56 23.630&-69 42 11.00& 12.82&  8.45&RSG   &    0.75&    3950&    4.90&& 0.06&3625&  4.80& 640&M2      \\
04562827-6940369&04 56 28.276&-69 40 36.95& 12.91&  8.43&RSG   &    0.75&    3900&    4.90&& 0.07&3675&  4.82& 630&M1.5    \\
05032723-6709129&05 03 27.238&-67 09 12.94& 13.19&  9.63&RSG+B &    0.75&    3950&    4.43&& 0.75&3825&  4.39& 360&K5-M0   \\
05040849-7014253&05 04 08.490&-70 14 25.30& 13.20&  9.47&RSG+B &    0.75&    4150&    4.55&& 0.75&3850&  4.46& 380&K5-M0   \\
05045253-7041578&05 04 52.532&-70 41 57.84& 13.31&  7.99&RSG   &    0.75&    3700&    5.02&& 0.11&3575&  4.99& 810&M2.5    \\
05045412-7033184&05 04 54.126&-70 33 18.49& 12.85&  8.40&RSG   &    0.75&    3900&    4.91&& 0.04&3625&  4.82& 650&M2      \\
05050732-7006123&05 05 07.320&-70 06 12.32& 12.72&  8.31&RSG+B &    0.75&    3900&    4.95&& 0.75&3725&  4.89& 670&M0      \\
05053350-7033469&05 05 33.502&-70 33 46.95& 12.98&  7.64&RSG   &    0.75&    3700&    5.16&& 0.25&3475&  5.11& 990&M4-4.5  \\
05053934-7038446&05 05 39.347&-70 38 44.65& 13.33&  9.37&RSG   &    0.75&    4100&    4.57&& 0.39&3850&  4.50& 400&K5-M0   \\
05092738-6831398&05 09 27.388&-68 31 39.89& 13.19&  8.82&RSG+B &    0.75&    3900&    4.74&& 0.75&3700&  4.68& 530&M1      \\
05121313-6804555&05 12 13.130&-68 04 55.50& 11.62&  7.32&RSG   &    0.75&    4100&    5.40&& 0.25&3725&  5.27&1030&M0      \\
05130492-6713314&05 13 04.925&-67 13 31.47& 13.00&  9.05&RSG+Be&    0.75&    3950&    4.67&& 0.75&3725&  4.60& 480&M0      \\
05133288-6921425&05 13 32.888&-69 21 42.51& 12.65&  8.08&RSG   &    0.75&    3900&    5.03&& 0.06&3650&  4.97& 760&M1.5    \\
05151642-6933065&05 15 16.426&-69 33 06.51& 12.62&  7.84&RSG   &    0.75&    3800&    5.11&& 0.75&3750&  5.09& 830&K5-M0   \\
05183040-6936218&05 18 30.406&-69 36 21.85& 13.14&  9.27&RSG   &    0.75&    4000&    4.60&& 0.33&3725&  4.50& 430&M0      \\
05185633-6756138&05 18 56.333&-67 56 13.81& 12.53&  7.52&RSG+Be&    0.75&    3850&    5.24&& 0.75&3600&  5.17& 990&M2      \\
05203947-6919310&05 20 39.470&-69 19 31.00& 13.28&  9.58&RSG+B &    0.75&    4100&    4.51&& 0.75&3775&  4.40& 370&K5-M0   \\
05205600-6528352&05 20 56.001&-65 28 35.21& 12.66&  8.11&RSG+B &    0.75&    3850&    5.01&& 0.75&3675&  4.96& 740&M1      \\
05230392-6704254&05 23 03.929&-67 04 25.48& 13.05&  8.67&RSG+B &    0.75&    3900&    4.81&& 0.75&3700&  4.74& 570&M1      \\
05241895-7026030&05 24 18.959&-70 26 03.08& 12.60&  8.28&RSG+B &    0.75&    3850&    4.94&& 0.75&3675&  4.89& 680&M1      \\
05254453-6616228&05 25 44.530&-66 16 22.80& 13.73&  9.51&RSG+B & \nodata& \nodata& \nodata&& 0.75&3750&  4.42& 380&M0      \\
05260034-7135488&05 26 00.342&-71 35 48.87& 12.86&  7.88&RSG+Be&    0.75&    3700&    5.06&& 0.75&3600&  5.03& 840&M2      \\
05270424-6726065&05 27 04.248&-67 26 06.59& 12.29&  8.23&RSG+B &    0.75&    3850&    4.96&& 0.75&3750&  4.93& 690&M0      \\
05272458-6653518&05 27 24.582&-66 53 51.84& 12.63&  8.41&RSG+B &    0.75&    3900&    4.91&& 0.75&3700&  4.84& 640&M1      \\
05272969-6714131&05 27 29.690&-67 14 13.10& 12.86&  7.97&RSG+B &    0.75&    3950&    5.09&& 0.75&3600&  4.99& 800&M2      \\
05273964-6909012&05 27 39.645&-69 09 01.21& 12.19&  7.97&RSG   &    0.75&    3900&    5.08&& 0.08&3675&  5.01& 790&M1      \\
05280004-6907424&05 28 00.040&-69 07 42.40& 13.11&  8.99&RSG   &    0.75&    4050&    4.72&& 0.22&3625&  4.58& 500&M2      \\
05281859-6907348&05 28 18.593&-69 07 34.80& 12.89&  8.31&RSG   &    0.75&    3750&    4.90&& 0.01&3650&  4.87& 680&M1.5    \\
05284914-6727256&05 28 49.140&-67 27 25.66& 13.11&  9.43&RSG+B &    0.75&    4050&    4.54&& 0.75&3825&  4.47& 390&K5-M0   \\
05285982-6717210&05 28 59.827&-67 17 21.03& 12.99&  9.16&RSG+B &    0.75&    4000&    4.63&& 0.75&3800&  4.57& 450&K5-M0   \\
05290550-6718175&05 29 05.500&-67 18 17.53& 12.85&  8.57&RSG   &    0.75&    3850&    4.82&& 0.12&3700&  4.76& 590&M1      \\
05291137-6628091&05 29 11.377&-66 28 09.14& 13.73&  9.84&RSG+B &    0.75&    4000&    4.36&& 0.75&3825&  4.31& 320&K5-M0   \\
05292757-6908502&05 29 27.570&-69 08 50.20& 12.29&  7.30&RSG+B &    0.75&    3850&    5.34&& 0.75&3550&  5.24&1100&M2.5    \\
05294618-6837024&05 29 46.184&-68 37 02.45& 13.67&  8.75&RSG+Be&    0.75&    3800&    4.74&& 0.75&3675&  4.70& 550&M1      \\
05294707-6714161&05 29 47.074&-67 14 16.10& 13.52&  9.63&RSG+B &    0.75&    3950&    4.43&& 0.75&3775&  4.38& 350&K5-M0   \\
05302094-6720054&05 30 20.940&-67 20 05.40& 12.79&  7.45&RSG+B &    0.75&    4050&    5.34&& 0.75&3725&  5.23& 990&M0      \\
05312426-6841336&05 31 24.266&-68 41 33.64& 13.07&  8.68&RSG+Be&    0.75&    3850&    4.78&& 0.75&3700&  4.74& 570&M1      \\
05312818-6703228&05 31 28.180&-67 03 22.80& 13.05&  8.81&RSG   & \nodata& \nodata& \nodata&& 0.21&3800&  4.70& 520&K5-M0   \\
05324407-6703406&05 32 44.079&-67 03 40.68& 13.36&  9.47&RSG   &    0.75&    3950&    4.50&& 0.41&3750&  4.41& 380&M0      \\
05324723-6621526&05 32 47.232&-66 21 52.67& 13.08&  8.93&RSG+B &    0.75&    3850&    4.68&& 0.75&3775&  4.66& 500&K5-M0   \\
05331113-6700380&05 33 11.138&-67 00 38.09& 13.39&  9.89&RSG   &    0.75&    4100&    4.38&& 0.56&3825&  4.26& 310&K5-M0   \\
05342683-6659583&05 34 26.830&-66 59 58.30& 12.61&  8.68&RSG+B &    0.75&    4050&    4.84&& 0.75&3775&  4.76& 560&M0      \\
05353296-6819323&05 35 32.967&-68 19 32.37& 13.49&  9.43&RSG   &    0.75&    3950&    4.52&& 0.39&3750&  4.43& 390&M0      \\
05355196-6922290&05 35 51.963&-69 22 29.03& 12.81&  8.45&RSG+B &    0.75&    3850&    4.87&& 0.75&3775&  4.85& 620&K5-M0   \\
05360634-6856407&05 36 06.347&-68 56 40.76& 12.93&  8.44&RSG+Be&    0.75&    3850&    4.88&& 0.75&3750&  4.85& 630&M0      \\
05374509-6920485&05 37 45.095&-69 20 48.59& 12.17&  7.72&RSG   &    0.75&    3600&    5.10&& 0.23&3525&  5.12& 970&M3.5    \\
05390424-6936039&05 39 04.247&-69 36 03.92& 13.34&  8.17&RSG+B &    1.43&    3750&    4.98&& 1.43&3700&  4.97& 750&M1      \\
05401638-6659303&05 40 16.380&-66 59 30.30& 13.96&  9.57&RSG+B & \nodata& \nodata& \nodata&& 0.75&3675&  4.37& 380&M1      \\
05402532-6915302&05 40 25.320&-69 15 30.20& 12.56&  8.78&RSG   & \nodata& \nodata& \nodata&& 0.18&3750&  4.72& 540&M0      \\
05402876-6915321&05 40 28.764&-69 15 32.10& 12.07&  8.13&RSG+B &    0.75&    3750&    4.97&& 0.75&3825&  4.99& 720&K5-M0   \\
05412153-6913228&05 41 21.531&-69 13 22.80& 13.08&  8.65&RSG   &    0.75&    3700&    4.75&& 0.10&3650&  4.76& 600&M1.5    \\
05415741-6912182&05 41 57.418&-69 12 18.22& 12.81&  8.74&RSG+B &    0.75&    3900&    4.78&& 0.75&3675&  4.71& 560&M1      \\
05420389-6913074&05 42 03.897&-69 13 07.41& 13.30&  8.74&RSG   &    0.75&    3850&    4.77&& 0.14&3675&  4.71& 560&M1      \\
05535411-6647126&05 53 54.110&-66 47 12.60& 12.89&  9.68&RSG   & \nodata& \nodata& \nodata&& 0.53&4000&  4.39& 330&K2-3    \\
\enddata
\tablenotetext{a} {From \citealt{ZaritskyLMC}.}
\tablenotetext{b} {From \citealt{2MASS}.}
\tablenotetext{c} {Typical uncertainty 150~K.}
\tablenotetext{d} {Adopted from photometry for the binaries.}
\tablenotetext{e}{Typical uncertainty 25~K.}
\end{deluxetable}

\begin{deluxetable}{l l l}
\tablecaption{\label{tab:fracs} Binary Fraction of RSGs}
\tablewidth{0pt}
\tablehead{
\colhead{Type of RSG Companion}
&\colhead{Percent}
&\colhead{Error}
}
\startdata
OB stars & 13.5 & +7.6/-6.7 \\
in eclipse & 3.6 & $\pm0.01$ \\
compact companions & 2.4 & $\pm0.01$ \\ \hline\hline
Total & 19.5 & +7.6/-6.7 \\
\enddata
\end{deluxetable}

\end{document}